\documentclass[11pt]{article}
\newcommand{\ba}{\begin{eqnarray}}
\newcommand{\ea}{\end{eqnarray}}
\newcommand{\nn}{\nonumber}

\newcommand{\ud}[3]{{ #1 }^{ #2 }_{\phantom{#2} #3 }}
\newcommand{\dud}[4]{{ #1 }_{ #2 \phantom{#3} #4 }^{\phantom{#2} #3 }}
\newcommand{\nt}{\nonumber\\}
\newcommand{\ul}{\underline}

\newcommand{\cN}{{\cal N}}

\usepackage{comment}
\usepackage{cite}

\def \II {I\hspace{-.1em}I\hspace{.1em}}
\def \IIA {\mbox{\II A\hspace{.2em}}}

\begin{document}
\begin{titlepage}

\begin{center}

\hfill UT-09-01
\vskip .5in

\textbf{\LARGE Lorentzian Lie (3-)algebra and
toroidal compactification of M/string theory}

\vskip .5in
Pei-Ming Ho$^\dagger$\footnote{
E-mail address: pmho@phys.ntu.edu.tw}, 
{\large
Yutaka Matsuo$^\ddagger$\footnote{
E-mail address: matsuo@phys.s.u-tokyo.ac.jp}
and 
Shotaro Shiba$^\ddagger$\footnote{
E-mail address: shiba@hep-th.phys.s.u-tokyo.ac.jp}
 }\\
\vskip 3mm
{\it\large
$^\dagger$
Department of Physics, Center for Theoretical Sciences \\
and Leung Center for Cosmology and Particle Astrophysics, \\
National Taiwan University, Taipei 10617, Taiwan,
R.O.C.}\\
\vskip 3mm
{\it\large
$^\ddagger$
Department of Physics, Faculty of Science, University of Tokyo,\\
Hongo 7-3-1, Bunkyo-ku, Tokyo 113-0033, Japan\\
\noindent{ \smallskip }\\
}
\vspace{60pt}
\end{center}
\begin{abstract}
We construct a class of Lie 3-algebras 
with an arbitrary number of pairs of generators 
with Lorentzian signature metric. 
Some examples are given and corresponding BLG models are studied.
We show that such a system in general describes 
supersymmetric massive vector multiplets
after the ghost fields are Higgsed.
Simple systems with nontrivial interaction 
are realized by infinite dimensional Lie 3-algebras
associated with the loop algebras.
The massive fields are then naturally identified with
the Kaluza-Klein modes by the toroidal compactification
triggered by the ghost fields. For example,
D$p$-brane with an (infinite dimensional) 
affine Lie algebra symmetry  $\hat g$
can be identified with D$(p+1)$-brane
with gauge symmetry $g$.
\end{abstract}

\end{titlepage}

\setcounter{footnote}{0}

\section{Introduction}

Recently, Bagger, Lambert~\cite{BL,BL2,BL3} and Gustavsson~\cite{G} 
constructed a three-dimensional supercomformal 
field theory as a multiple-M2-brane
world-volume theory in M-theory. 
This BLG model is characteristic of the novel feature 
that the gauge symmetry is based on a Lie 3-algebra, 
and thus various studies on this algebra have been undertaken~\cite{HHM,Lie3algebra}.
For the BLG model to work, the Lie 3-algebra needs to satisfy
the fundamental identity (a generalization of Jabobi identity). 
If the positivity of the invariant metric is also imposed 
to avoid ghosts, 
the only non-trivial example of finite dimensional 3-algebra is
${\cal A}_4$~\cite{K} and its direct sums.

If we relax the condition on dimensionality, 
Nambu-Poisson brackets give realizations of 
infinite dimensional Lie 3-algebra~\cite{HM,HIMS}. 
The BLG model with this algebra realizes the world-volume theory
of M5-branes in the $C$-field background on
a 3-manifold where Nambu-Poisson bracket can be defined.

Similarly, 
when the requirement of a positive definite metric is given up, 
we also found physically meaningful models. 
Among the various examples, a Lie 3-algebra with 
a negative-norm generator was constructed
and was referred to as {\em Lorentzian} Lie 3-algebra.
\footnote{The Lie 3-algebra 
with zero-norm generators was also studied~\cite{CHMS}
to construct M2-brane model which produces the correct
entropy $O(N^{3/2})$ in large $N$ limit.
It was suggested that we need 3-algebra
instead of Lie algebra to have such scaling.}
The corresponding BLG model has ghosts, 
but they can be completely decoupled. 
It was realized that the inclusion of the Lorentzian generators 
is associated with the compactification of 
a spatial dimension, 
and this Lorentzian model reproduces the multiple-
D2-brane world-volume theory in type \IIA string theory. 

In this paper, we study some generalizations of such Lorentzian
3-algebras for which ghost fields can still be decoupled. 
Such algebras have been considered extensively 
by de Medeiros {\it et. al}~\cite{dMFM} when 
the number of Lorentzian pairs is two.
Here we present more straightforward and explicit analysis
in terms of the structure constants. 
We find it fruitful to consider generalizations 
with more Lorentzian pairs, 
as it gives us insight about how to 
circumvent the strict constraints from fundamental identities.
We also study the BLG model associated with such 3-algebras. 
Our construction includes an interesting example 
which contains the massive Kaluza-Klein towers 
associated with additional compactified dimensions. 
This seems to be consistent
with our expectation that adding Lorentzian pairs 
corresponds to additional compactifications. 
A typical feature of the generalized Lorentzian 3-algebra 
is indeed that we have a massive spectrum with $\cN=8$ SUSY 
in the BLG model, and we need an infinite dimensional 
realization to have nontrivial interacting models.

Our observation of the relation between the D-brane
system with Lorentzian gauge symmetry 
and higher dimensional branes is not restricted to 
the context of BLG models.
In fact, most of the examples considered here can be 
directly analyzed in the context of a Yang-Mills system 
whose gauge symmetry has Lorentzian signiture. 
It was known that in some brane configurations
(see for example \cite{DeWolfe:1998pr}) we have to treat such
an infinite dimensional gauge symmetry on D-branes.
It was generally expected that the appearance of infinite dimensional
symmetry should be related to closed string modes
in a compactified space.  However, the explicit analysis was not made
because the Higgs mechanism which implement Kaluza-Klein mass
was not known.  Similar infinite dimensional symmetries were also
studied in various contexts \cite{Lorentzian} 
in string/M theory and we hope that
our method gives a simple direct interpretation to such
systems. 

This paper is organized as follows.
In \S \ref{s:L-rev}, we first review the Lorentzian BLG model
\cite{GMR,Benvenuti:2008bt,HIM}.
We describe the typical structure of the 3-algebra
for which the removal of the ghost field 
\cite{Bandres:2008kj,Gomis:2008be}
is possible.
In \S\ref{s:Lie3}, we give a detailed study
of the constraint from the fundamental
identity.  Such study for two Lorentzian pairs was made 
in \cite{dMFM} but we generalize their result by considering
an arbitrary number of Lorentzian pairs. We use a strategy to
 analyze the  constraint for the structure constants directly. Although
we do not claim that we could classify all possible algebras, 
we find a class of interesting 3-algebras through such analysis, 
with potential applications to string/M theory.
We note that the many 3-algebras which we found
can be realized by Lorentzian extension 
\cite{GMR,Benvenuti:2008bt,HIM} of Lorentzian Lie algebras.
It enables us to analyze some of the Lorentzian BLG models 
through gauge theories with Lorentzian Lie algebra symmetry.
As we noted, such D-brane system is by itself an interesting
object to study. 
In \S\ref{sec:2A}, we derive the BLG model
associated with the simplest Lie 3-algebra 
with more than one Lorentzian pairs.
We demonstrate that such system typically has massive vector fields
where each gauge field absorbs two degrees of freedom 
from scalar fields.
In \S\ref{D2D3}, 
we construct the BLG model (or super Yang-Mills theory) 
based on loop algebras which are the simplest nontrivial
examples of generalized Lorentzian Lie (3-)algebra.
Finally we comment that the description of M5-brane \cite{HM, HIMS}
can be also regarded as the typical example of the compactification
through the Lorentzian 3-algebra.

\section{Lorentzian BLG model}
\label{s:L-rev}

In this section, we review the 
basic features of the {\em Lorentzian} BLG model
~\cite{HIM,GMR,Benvenuti:2008bt}.
The original BLG action for multiple M2-branes is 
\ba
S&=&T_2\int d^3x\, L 
=T_2\int d^3x\, (L_X+L_\Psi+L_{int}+L_{pot}+L_{CS}), \label{S} \\
L_X&=&-\frac{1}{2}\langle D_\mu X^I , D^\mu X^I \rangle, \label{LX}\\
L_\Psi&=&\frac{i}{2}\langle \bar\Psi , \Gamma^\mu D_\mu \Psi \rangle, 
\label{LPsi} \\
L_{int}&=&\frac{i}{4}\langle\bar\Psi , \Gamma_{IJ}[X^I,X^J,\Psi] \rangle, 
\label{Lint} \\
L_{pot}&=&
-\frac{1}{12} \langle [X^I, X^J, X^K], [X^I, X^J, X^K]\rangle, 
\label{Lpot} \\
L_{CS}&=& \frac{1}{2}f^{ABCD} A_{AB}\wedge d A_{CD}+
\frac{1}{3} {f^{CDA}}_G f^{EFGB} A_{AB}\wedge
A_{CD}\wedge  A_{EF}\,, \label{LCS}
\ea
where $T_2$ is the M2-brane tension. 
The indices $\mu=0,1,2$ specify the longitudinal directions of M2-branes; 
$I,J,K=3,\cdots,10$ the transverse directions. 
The indices $A,B,C,\cdots$ denote components 
of Lie 3-algebra generators. 
The covariant derivative is
\ba \label{eq:cov}
 (D_\mu \Phi(x))_A =\partial_\mu \Phi_A-{f^{CDB}}_A A_{\mu CD}(x) \Phi_B
\ea
for $\Phi=X^I,\Psi$.


In order to define the BLG model action, 
the Lie 3-bracket
\ba
[T^A, T^B, T^C]={f^{ABC}}_D T^D
\ea
for a Lie 3-algebra must satisfy the following constraints:
\begin{itemize}
\item Tri-linearity
\item Skew symmetry
\item Fundamental identity
\ba\label{e:FI}
{f^{ABC}}_F{f^{FDE}}_G+
{f^{ABD}}_F{f^{CFE}}_G+
{f^{ABE}}_F{f^{CDF}}_G=
{f^{CDE}}_F {f^{ABF}}_G
\ea
\item Invariant metric $\langle T^A, T^B\rangle=h^{AB}$:
\ba
{f^{ABC}}_E h^{ED}+{f^{ABD}}_E h^{CE}=0\,.
\ea
\end{itemize}

The simplest Lorentzian Lie 3-algebra was defined as follows.
Let $\mathcal G$ be a given Lie algebra. 
We denote its generators as $T^i$, 
structure constants $\ud f{ij}k$, and Killing form $h^{ij}$.
Now we define a Lie 3-algebra whose generators 
are $T^A=\{u,v,T^i\}$ such that
\ba \label{L1}
&&
{[v,T^A,T^B]}=0,\quad
{[u,T^i,T^j]}={\ud f{ij}k}T^k,\quad
{[T^i,T^j,T^k]}=-h^{kl}{\ud f{ij}l}v,
\nt &&
\langle u,v \rangle = 1,\quad
\langle T^i,T^j \rangle = h^{ij},\quad
\mbox{otherwise}=0.
\ea
This 3-algebra satisfies the fundamental identities 
and the requirement of invariant metric, 
so we can use it as the gauge symmetry of BLG model.
Since this algebra has a negative-norm generator 
$u-\alpha v$ (for $\alpha>0$),
BLG model has a ghost field. 
The mode expansion of the Langrangian becomes 
(up to total derivatives)
\ba
L&=&\left\langle
-\frac12 (\hat D_\mu \hat X^I-A_\mu'X_i^I)^2
+\frac{i}2 \bar{\hat\Psi}\Gamma^\mu \hat D_\mu \hat\Psi
+\frac{i}2 \bar \Psi_u\Gamma^\mu A_\mu' \hat \Psi
\right.\nt&&\quad
+\frac{i}2 \bar{\hat\Psi}\Gamma_{IJ}X_u^I[\hat X^J,\hat\Psi]
+\frac14(X_u^K)^2[\hat X^I,\hat X^J]^2
-\frac12(X_u^I[\hat X^I,\hat X^J])^2
\nt&&\quad\left.
+\frac12 \epsilon^{\mu\nu\lambda}\hat F_{\mu\nu} A'_\lambda \right\rangle
+L_{gh}, \\
L_{gh}&=& -\left\langle 
\partial_\mu X_u^I A'_\mu \hat X^I
+(\partial_\mu X_u^I)(\partial_\mu X_v^I)
-\frac{i}2\bar\Psi_v \Gamma^\mu\partial_\mu \Psi_u
\right\rangle,
\ea
where 
\ba
\hat D_\mu \Phi:=\partial_\mu \hat \Phi-[\hat A_\mu,\hat \Phi],\quad
\hat F_{\mu\nu}:=\partial_\mu \hat A_\nu - \partial_\nu \hat A_\mu
-[\hat A_\mu,\hat A_\nu]
\ea
for $\Phi=X^I,\Psi$.
As we see, fortunately, the ghost fields decouple, 
that is, they act only as Langrange multipliers. 
Their equations of motions are
\ba\label{g-f}
\partial_\mu^2 X_u^I=0,\quad
\Gamma^\mu\partial_\mu \Psi_u=0,
\ea
and we can set 
\ba \label{setVEV}
X_u^I=\lambda^I:=\lambda\delta_{10}^I,\qquad
\Psi_u^I=0
\ea
without breaking any supersymmetry or gauge symmetry
\cite{HIM}.  
This is motivated by the Higgs mechanism in BLG model 
first considered in \cite{MP}.
The Lagrangian becomes, after integration over $A'$,
\ba\label{eLagYM}
L=-\frac12 (\hat D_\mu \hat X^I)^2
+\frac{i}2 \bar{\hat\Psi} \Gamma^\mu \hat D_\mu \hat\Psi
+\frac{\lambda^2}4 [\hat X^I,\hat X^J]^2
+\frac{i\lambda}2 \bar{\hat\Psi} \Gamma_I [X^I, \hat \Psi]
-\frac1{4\lambda^2} \hat F_{\mu\nu}^2,
\ea
where $I,J=3,\cdots,9$.
This can be regarded as D2-branes theory in type \IIA string theory
which is the compactification of M-theory on a circle.

The origin of the decoupling of the ghost fields comes from the
specific way that Lorentzian generators appear in the 3-algebra.
Namely, the generator $v$ is the center of the 3-algebra and 
$u$ is not produced in any 3-commutators.  This property
ensures that the system is invariant under the
translation of the scalar fields $X^I_u$.
The decoupling of the ghost fields can be made more rigorous
\cite{Bandres:2008kj, Gomis:2008be} 
by gauging this global symmetry. Namely by adding
extra gauge fields $C_\mu, \chi$ through
\ba
\label{Xugauge}
L_{new}=-\bar \Psi_u \chi +\partial^\mu X^I_u C_\mu^I\,,
\ea
we have an extra gauge symmetry:
\ba
\delta X^I_v=\Lambda^I,\qquad
\delta C^I_\mu =\partial_\mu \Lambda^I,\qquad
\delta \Psi_v =\eta,\qquad
\delta\chi=i\Gamma^\mu\partial_\mu \eta\,.
\ea
It enable us to put $X^I_v=\Psi_v=0$. The equations of
motion by variation of $C^I_\mu,\chi$ give 
the assignment (\ref{setVEV}) correctly.

Another important feature of the Lorentzian BLG model
is that the assignment of VEV to $X^I_u$ triggers
the compactification of 11 dimensional M-theory
to 10 dimensional type \IIA theory.  
The compactification radius of M-direction is given by
\cite{HIM}
\ba\label{lamR}
\lambda=2\pi R\,.
\ea
For various aspects of the Lorentzian model, see for example \cite{rLBLG}.

\section{Analysis of Lie 3-algebra with two or more 
negative-norm generators}
\label{s:Lie3}

In the following, we consider some generalizations
of the Lorentzian 3-algebra invented in \cite{HIM,GMR,Benvenuti:2008bt}
by adding pairs of generators with
Lorentzian metric.
Positive-norm generators are denoted as $e^i$ ($i=1,\cdots, N$),  
and Lorentzian pairs as $u_a, v_a$ ($a,b=1,\cdots, M$). 
We assume that the invariant metric for them is given
by the following simple form
\ba\label{e:metric}
\langle e^i, e^j\rangle= \delta^{ij}\,,\quad
\langle u_a, v_b\rangle=\delta_{ab}\, .
\ea
In terms of the four-tensor defined by 
\ba
f^{ABCD}:={f^{ABC}}_E h^{ED}\,,
\ea
the invariance of the metric and the skew symmetry of 
the structure constant imply that
the condition that this 4-tensor
is anti-symmetric with respect to all indices.

We also assume that the generators $v_a$ are 
in the center of the 3-algebra.
This condition is necessary to apply the Higgs mechanism
to get rid of the ghost fields as we have reviewed.
In terms of the 4-tensor this condition
is written as
\ba\label{e:str}
f^{v_a BCD}=0
\ea
for arbitrary $B,C,D$.  Therefore the index in the 4-tensor
is limited to $e^i$ and $u_a$.  For the simplicity of the notation,
we write $i$ for $e^i$ and $a$ for $u_a$ for indices of the 4-tensor,
for example $f^{ijab}:=f^{e^i e^j u_a u_b}$ and so on.

We note that there is some freedom in 
the choice of basis when keeping
the metric (\ref{e:metric}) and the form of 4-tensor (\ref{e:str})
invariant:
\ba
\tilde e^i= O^i_j e^j + P^i_a v^a,\quad
\tilde u^a= Q^{a}_i e^i + R^a_b u^b + S^a_b v^b,\quad
\tilde v^a= ((R^t)^{-1})^a_b v^b,\label{change}
\ea
where
\ba
O^t O=1,\quad
Q=-R  P^t O,\quad
R^{-1} S+ (R^{-1} S)^t=- P^tP\,.
\ea
The matrices $O$ and $R$ describe the usual rotations of the basis.
The matrix $P$ describes the mixing of the Lorentzian
generators $u_a, v_a$ with $e^i$.

We introduce some notation for the 4-tensor,
\ba\label{str1}
f^{ijkl}=F^{ijkl},\ 
 f^{a ijk}=f^{ijk}_a,\ 
 f^{a b ij}=J_{ab}^{ij},\ 
 f^{a b c i}=K_{abc}^i,\ 
 f^{abcd}=L_{abcd}\,.
\ea
We rewrite the fundamental identity in terms of this notation
below in \S\ref{FIs}.

There are a few comments which can be made without detailed analysis:
\begin{itemize}
\item For lower $M$ ({\it i.e.} smaller number of Lorentzian pairs $(u_a, v_a)$),
some components of the structure constants (\ref{str1})
vanish identically due to the anti-symmetry of indices.  For example,
for $M=1$, we need to put $J^{ij}_{ab}= K^{i}_{abc}=L_{abcd}=0$.
For $M=2$, one may put $J^{ij}_{ab}$ nonvanishing but 
we have to keep $K^{i}_{abc}=L_{abcd}=0$ and so on.

\item In the fundamental identity (\ref{f1}--\ref{f23}),
there is no constraint on $L_{abcd}$.  It comes from the fact
that the contraction with respect to Lorentzian indices automatically
vanishes due to the restriction of the structure constant (\ref{e:str}).
So it can take arbitrary value for $M\geq 4$.
This term, however, is not physically relevant in BLG model,
since they appear only in the interaction terms of the ghost
fields which will be erased after Higgs mechanism.

\item A constraint for $F^{ijkl}$ (\ref{f1}) is identical to
the fundamental identity of a 3-algebra with the structure constant $F^{ijkl}$. 
So if we assume positive definite metric for $e^i$, 
it automatically implies that $F^{ijkl}$ is
proportional to $\epsilon_{ijkl}$ or its direct sums \cite{no-go}.

\item By a change of basis (\ref{change}),
various components of the structure constants (\ref{str1}) mix.  
For example, if we put
$O=R=1$ for simplicity and keep only the matrix $P$
nontrivial (which implies $S=-\frac{1}{2} P^t P$),
the structure constant in terms of the new basis $\{\tilde e^i$,
$\tilde u^a$, $\tilde v^a\}$ are given as
\ba
\tilde F^{ijkl}&=&F^{ijkl}, \label{r1}\\
\tilde f^{jkl}_a&=& f^{jkl}_a +P^i_a F^{ijkl}, \label{r2}\\
\tilde J^{ij}_{ab}& = & J^{ij}_{ab}+P^k_a f^{ijk}_b-P^k_b f^{ijk}_a
+ F^{ijkl} P^k_a P^l_b, \label{r3}\\
\tilde K^i_{abc}&=& K^i_{abc}+P^j_a J^{ij}_{bc}-P^j_b J^{ij}_{ac}
+ P^j_c J^{ij}_{ab}, \nn\\
&&+f^{ikl}_c P^k_a P^l_b-f^{ikl}_b P^k_a P^l_c + f^{ikl}_a P^k_b P^l_c
+P^j_a P^k_b P^l_c F^{ijkl}\,. \label{r4}
\ea
We will find that many solutions of the fundamental identities
can indeed  be identified with well-known 3-algebra
after such redefinition of basis.  In this sense, 
the classification of the Lorentzian 3-algebra has a
character of cohomology, namely only solutions which can 
not reduce to known examples after all changes
of basis give rise to physically new system.
\end{itemize}

In the following, we give a somewhat technical analysis 
of the fundamental identity (\ref{e:FI}).
Solutions which we found are summarized in \S\ref{s:sum}.
We do not claim that our analysis exhausts all the possible
solutions.  But as we will see in the later sections,
they play an important physical role in string/M theory
compactification.

\subsection{Fundamental identities}
\label{FIs}

We rewrite the fundamental identity (\ref{e:FI})
in the notation (\ref{str1}):
\ba
&& F^{ijkn}F^{nlmp}+F^{ijln}F^{knmp}
+F^{ijmn}F^{klnp}-F^{klmn}F^{ijnp}=0, \label{f1}\\
&& F^{ijkn}f^{nlm}_a +
F^{ijln}f^{knm}_a+F^{ijmn}f^{kln}_a-F^{klmn}f^{ijn}_a=0, 
\label{f2}\\
&&f^{ijn}_aF^{nklm}+f^{ikn}_aF^{jnlm}+f^{iln}_aF^{jknm}
-f^{inm}_aF^{jkln}=0, \label{f3}\\
&& (f^{ijn}_a f^{nkl}_b+f^{ikn}_af^{jnl}_b
+f^{iln}_af^{jkn}_b) +F^{jkln}J^{in}_{ab}=0, \label{f4}\\
&& J^{im}_{ab}F^{mjkl}+J^{jm}_{ab}F^{imkl}
+J^{km}_{ab}F^{ijml}+J^{lm}_{ab}F^{ijkm}=0, \label{f5}\\
&&(J^{im}_{ab} f^{mjk}_c+J^{jm}_{ab}f^{imk}_c+J^{km}_{ab}f^{ijm}_c)
-F^{ijkm} K^m_{abc}=0, \label{f6}\\
&&F^{ijkn}J^{nl}_{ab}-F^{ijln}J^{nk}_{ab}
-f^{ijn}_af^{nkl}_b+f^{ijn}_b f^{nkl}_a
=0, \label{f7}\\
&& (J_{ab}^{im}f_{c}^{mjk}-J_{ac}^{im} f_{b}^{mjk})
+(f^{ijm}_aJ^{mk}_{bc}-f^{ikm}_aJ^{mj}_{bc})=0, \label{f9}\\
&&-K_{abc}^l f_{d}^{lij}+K_{abd}^lf_{c}^{lij}
+ J_{ab}^{il}J_{cd}^{lj} - J_{cd}^{il}J_{ab}^{lj}
=0, \label{f11}\\
&& (f_a^{ikm}J_{bc}^{mi}+f_b^{jkm}J_{ca}^{mi}
+f_c^{jkm}J_{ab}^{mi})
+K^{m}_{abc}F^{jkim}=0, \label{f13}\\
&& (J_{ab}^{jl}J_{cd}^{li}+J_{ad}^{jl}J_{bc}^{li}-J_{ac}^{jl}J_{bd}^{li})
-f^{jil}_cK^{l}_{abd}=0, \label{f15}\\
&& -J_{ab}^{ki}K_{cde}^k-J_{be}^{ki}K_{acd}^k
+J_{ae}^{ki}K_{bcd}^k+J_{cd}^{ki}K_{abe}^k=0, \label{f17}\\
&& f_a^{ijl}K_{bcd}^l-f_b^{ijl}K_{acd}^l
+f_c^{ijl}K_{abd}^l-f^{ijl}_dK_{abc}^l=0, \label{f19}\\
&& K_{abc}^iK_{def}^i-K_{ade}^iK_{bcf}^i
+K_{acf}^iK_{bde}^i-K_{abf}^iK_{cde}^i=0. \label{f23}
\ea

\subsection{Lorentzian extension of Nambu bracket}
\label{Fneq0}

Let us examine the case with $F^{ijkl}\neq 0$ first.
As we already mentioned, eq.(\ref{f1}) implies that
$F^{ijkl}\propto \epsilon_{ijkl}$ and its direct sum.
So without losing generality, one may assume 
$N=4$ and $F^{ijkl}=\epsilon_{ijkl}$
for the terms which include nontrivial 
contraction with $F^{ijkl}$.

Suppose $f_{a}^{ijk}\neq 0$ for some $a$.  Then
by the skew-symmetry of indices they can be written as
$f^{ijk}_a=\epsilon_{ijkl} P_l^a$
for some $P_l^a$.  This expression actually
solves (\ref{f2},\ref{f3}).
However, this form of $f^{ijk}_a$ is 
exactly the same as the right hand side of (\ref{r2}).
It implies that such $f^{ijk}_a$ can be set to zero 
by a redefinition of basis.

Therefore, at least when the 3-algebra is finite dimensional,
it is impossible to construct Lorentzian algebra
with nontrivial $F^{ijkl}\neq 0$.  The situation is totally different
if the 3-algebra is infinite dimensional \cite{HM,HIMS}
which is related to the description of M5-brane
(for the various aspects of M5-brane in BLG context, see also \cite{rM5} for example).
The realization of the three-algebra was given as follows.
We take $\mathcal{N}$ as a compact three dimensional manifold
where Nambu-Poisson bracket
\cite{Nambu},
\ba\label{eNP}
\left\{
f_1, f_2, f_3
\right\}=\sum_{a,b,c}\epsilon_{abc}\, \partial_a f_1 \partial_b f_2 \partial_c f_3
\ea
is well defined.  Namely $\mathcal{N}$ is covered by the local coordinate patches
where the coordinate transformation between the two patches
keeps the 3-bracket (\ref{eNP}) invariant.  The simplest examples
are $T^3$ and $S^3$ \cite{HM,CHMS}.
If we take $\chi^i(y)$ as the basis of $\mathcal{H}$:
the Hilbert space which consists of functions which are globaly well-defined
on $\mathcal{N}$, and one can choose a basis mutually orthonormal with respect to
the inner product,
\ba
\langle \chi^i, \chi^j\rangle := 
\int_{\mathcal{N}} d^3 y\, \chi^i(y) \chi^j(y)=\delta^{ij}\,.
\label{MIP}
\ea
It is known that the structure constant
\ba\label{M5F}
F^{ijkl}=\langle \left\{\chi^i, \chi^j, \chi^k\right\},\chi^l\rangle
\ea
satisfies the fundamental identity (\ref{f1}).


We are going to show that it is possible to extend this 3-algebra 
with the additional generators with
the Lorentzian signature.  
For simplicity, we consider the case
$\mathcal{N}=T^3$.  The Hilbert space $\mathcal{H}$ is spanned by
the periodic functions on $T^3$.  If we write the flat coordinates
on $T^3$ as $y^a$ ($a=1,2,3$), where the periodicity is imposed
as $y^a\sim y^a+ p^a$, and $p^a\in \mathbf{Z}$.
The basis of $\mathcal{H}$ is then given by
\ba
\chi^{\vec n}(y) := e^{2\pi i n_a y^a}\,,\quad
\vec n \in \mathbf{Z}^3\,,
\ea
with the invariant metric and the structure constant:
\ba
&& \langle \chi^{\vec n}, \chi^{\vec m}\rangle =\delta(\vec n+\vec m)\,,
\label{Tmet}\\
&& F^{\vec n\vec m\vec l\vec p}=(2\pi i)^3 \epsilon_{abc} n^a m^b l^c
\delta(\vec n +\vec m+\vec l+\vec p)\label{M5Fa}\,.
\ea

The idea to extend the 3-algebra is to introduce the functions
which are {\em not} well-defined on $T^3$ but the Nambu bracket
among $\mathcal{H}$ and these generators remains in $\mathcal{H}$.
For $T^3$, such generators are given by the functions $u_a = y^a$.
The fundamental identity for the Nambu-bracket comes from the
definition of derivative and it does not matter whether or not the functions
in the bracket is well-defined globally.  Therefore even if we include
extra generators the analog of fundamental identity holds.
More explicitly we define the extra structure constants as
\ba
f^{\vec n\vec m\vec l}_a&:=& \langle\left\{
u^a, \chi^{\vec n},\chi^{\vec m}\right\},\chi^{\vec l}\rangle
=(2\pi i)^2 \epsilon_{abc} n^b m^c \delta(\vec n +\vec m +\vec l),
\label{M5f}\\
J^{\vec n \vec m}_{ab}&:=& \langle\left\{
u^a, u^b ,\chi^{\vec n}\right\},\chi^{\vec m}\rangle
= (2\pi i) \epsilon_{abc} n^c \delta(\vec n + \vec m),
\label{M5J}\\
K^{\vec n}_{abc}&:=& \langle\left\{u^a, u^b, u^c\right\},\chi^{\vec n}\rangle
=\epsilon_{abc} \delta(\vec n).
\label{M5K}
\ea
It is not difficult to demonstrate explicitly that
they satisfy all the fundamental identities (\ref{f1}--\ref{f23}).

We have to be careful in the treatment of the new generators.
For example, the inner product (\ref{MIP}) is not well-defined
if the function is not globally well-defined on $\mathcal{N}$.
The fact that the structure constants 
(\ref{M5Fa}--\ref{M5K}) satisfies the fundamental identities
(\ref{f1}--\ref{f23}) implies that we can define the inner product
{\em abstractly} as (\ref{e:metric}).  Namely we introduce extra
generators $v_a$ ($a=1,2,3$) and define 
\ba
\langle u_a, v_b\rangle =\delta_{ab},\quad
\langle u_a, \chi^{\vec n}\rangle =
\langle v_a, \chi^{\vec n}\rangle = 
\langle u_a, u_b\rangle = 
\langle v_a, v_b\rangle = 
0
\ea
while keeping  (\ref{Tmet}).

We also need to be careful in the definition of the three-bracket
itself.  The naive Nambu bracket needs to be modified
to make the structure constant $F^{ABCD}$ totally anti-symmetric
in all four indices.  This condition is broken in the original Nambu
bracket after the introduction of the extra generators $u^a$.
We have to come back to our original definition of 3-algebra
where this symmetry is manifest.  This implies the following redefinition 
of the 3-algebra:
\ba
\left[\chi^{\vec n},  \chi^{\vec m}, \chi^{\vec l}\right]&=& 
{F^{\vec n\vec m\vec l}}_{\vec p}\chi^{\vec p} -f^{\vec n\vec m\vec l}_a v^a,
\label{3new1}\\
\left[ u^a, \chi^{\vec n}, \chi^{\vec m}\right]&=& {f^{\vec n\vec m}_a}_{\vec l}
 \chi^{\vec l} + J^{\vec n\vec m}_{ab} v^b, \\
\left[u^a, u^b, \chi^{\vec n}\right]&=& {J^{\vec n}_{ab}}_{\vec m} \chi^{\vec m}
 -K^{\vec n}_{abc} v^c, \\
\left[u^a, u^b,u^c\right]&=& K_{abc\vec n}\chi^{\vec n}.\label{3new4}
\ea
This 3-algebra may be regarded as the ``central extension" 
of the Nambu-Poisson bracket.  The additional factors which are
proportional to $v^a$ on the right hand side is necessary to
make the metric invariant.  One might worry if the fundamental identity
may be violated by the redefinition of the algebra.  In this example,
fortunately this turns out not to be true.  So we have a consistent 3-algebra
with Lorentzian signiture.
It may be useful to repeat our emphasis that,
although $u^a$ was originally defined through ill-defined
function $y^a$, we have to neglect this fact to define the metric 
and the 3-algebra. 

While the 3-algebra (\ref{3new1}--\ref{3new4}) is new, we will see later 
in \S\ref{sM5}
that the BLG model based on it turns out to be the same as
the M5 models defined in \cite{HM, HIM, HIMS}
although it was not noticed explicitly.  A glimpse of this fact
appeared in \S7 in \cite{HIM} where a subalgebra of
(\ref{3new1}--\ref{3new4}) appeared and the relation
with the Lorentzian BLG model and M5 model was discussed. 
We will give more comments on this issue later in \S\ref{sM5}.


It is straightforward to obtain similar Lorentzian extensions of Nambu-Poisson
type Lie 3-algebras defined on different manifolds $\mathcal{N}$
such as $S^3$ and $S^2\times S^1$.
So far, the only nontrivial Lie 3-algebra with positive definite metric are
${\cal A}_4$ and the Nambu-Poisson type 3-algebras. 
The examples
we consider here would exhaust the Lorentzian extensions which can
be obtained from them.

\subsection{Constraints from the fundamental identities for $F^{ijkl} = 0$}

In the following, we restrict ourselves to
the case  $F^{ijkl}=0$.
The fundamental identities (\ref{f1}--\ref{f23})
are now simplified to be the following:
\ba
&& f^{ni(j}_a f^{kl)n}_b = 0, \label{f4'}\\
&& f_{(a}^{ijm} f_{b)}^{mkl} = 0, \label{f4-1}\\
&& f_a^{ijm} f_b^{mkl} + f_a^{kim} f_b^{mjl} + f_b^{jkm} f_a^{mil} 
= 0,
\label{f4-2}\\
&& J_{ab}^{l(i} f_c^{jk)l} = 0, \label{f6/f10}\\
&& f_{(a}^{ijk}J_{bc)}^{kl} = 0, \label{f13/f18}\\
&& J_{a(b}^{il}f_{c)}^{ljk} + J_{bc}^{l(j} f_{a}^{k)il} = 0, \label{f9'}\\
&&2K_{ab(c}^k f_{d)}^{kij}=
J_{ab}^{ik}J_{cd}^{kj}-J_{cd}^{ik}J_{ab}^{kj}, \label{f11-1}\\
&& f_a^{ijk} K^{k}_{bcd} = 3J_{a(b}^{ik}J_{cd)}^{kj}, \label{f15/f20}\\
&& 3K_{ab(c}^i J_{de)}^{ij} = K_{cde}^i J_{ab}^{ij}, \label{f17'}\\
&& J_{a(b}^{ij} K^j_{cde)} = 0, \label{f17-1}\\
&& f_{(a}^{ijk}K_{bcd)}^k = 0, \label{f19'}\\
&& 3K_{ab(c}^i K_{de)f}^i = K_{cde}^i K_{abf}^i \,. \label{f23'}
\ea
In the above, we used the notation that all indices in parentheses 
are fully antisymmetrized. 
For instance, 
\ba
A_{a(b}B_{cd)e} := 
\frac{1}{6}\left(
A_{ab}B_{cde}+A_{ac}B_{dbe}+A_{ad}B_{bce}
-A_{ab}B_{dce}-A_{ac}B_{bde}-A_{ad}B_{cbe}\right). 
\ea

The constraints above are not all independent. 
We can use (\ref{f4-2}) alone to derive (\ref{f4'}) and (\ref{f4-1}) 
as follows. 
Taking (\ref{f4-2}) and replacing the indices as
$(ijk) \rightarrow (jki)$ and $(ab) \rightarrow (ba)$ 
and subtracting the derived equation from (\ref{f4-2}), 
we get (\ref{f4-1}). 
It is also obvious that (\ref{f4-1}) and (\ref{f4-2}) implies (\ref{f4'}). 

Similarly, (\ref{f9'}) can be easily derived from 
(\ref{f6/f10}) and (\ref{f13/f18}).

\subsection{Solutions}

In this subsection, we try to solve
the fundamental identities displayed above
and find a class of solutions.

First, 
a solution for (\ref{f4'}) is to use a direct sum of Lie algebras 
$g=g_1\oplus\cdots \oplus g_n$,  
We divide the values of indices into
$n$ blocks $I=I_1\cup \cdots \cup I_n$ and
let
\ba\label{e:f}
f^{ijk}_a=\gamma_a^\alpha f^{ijk}_\alpha,
\ea
where $f^{ijk}_\alpha$ is defined by 
\ba
f^{ijk}_\alpha=\left\{
\begin{array}{ll}
f^{ijk}_{g_\alpha} \quad&i,j,k\in I_\alpha, \\
0 & \mbox{otherwise}.
\end{array}
\right.
\ea
Here $f^{ijk}_{g_\alpha}$ is the structure constant 
for $g_\alpha$ while $\gamma^\alpha_a$ is a real number.

Note that the number $n$ does not have to equal $M$. 
It is possible to have some of the sets $I_a$ empty. 
An example has $g = g_1$ and 
all $I_{a\neq 1}$ empty. 
In this case, for $\gamma^{\alpha}_a = \delta^{\alpha}_a$,
we have $f^{ijk}_1 = f^{ijk}_{g_1}$ and 
$f^{ijk}_a = 0$ for all $a\neq 1$.

If all the other components of the 3-algebra structure constant vanish,
one obtains from (\ref{e:f}) a set of solutions to the fundamental identity.
The BLG model for this 3-algebra is not new, however.  
For each range of index,
say $I_\alpha$, we have
\ba
[e^i,e^j,e^k]= -\sum_a \gamma^\alpha_a v^a\,,\quad
[u^a, e^i, e^j]=\sum_k \gamma^\alpha_a f^{ijk}_\alpha e^k\,.
\ea
By a suitable rotation (\ref{change}) with 
\ba
v'^1=\sum_a \gamma^\alpha_a v^a,
\ea
we always have
\ba
[e^i,e^j,e^k]= - v'^a\,,\quad
[u'^a, e^i, e^j]=\delta_{a1} \sum_k  f^{ijk}_\alpha e^k\,.
\ea
Therefore it is reduced to the standard Lorentzian Lie 3-algebra 
for $M = 1$ after the restriction of indices to $I_\alpha$.

In order to obtain something new,
we have to allow other coefficients to be nonzero.

The simplest class of solutions can be found
when $f^{ijk}_a=0$ for $i,j,k\in I_a$.
In this case, for this range $I_a$, arbitrary
anti-symmetric matrix $J^{ij}$ ($i,j\in I_a$) 
solves the constraints (this case is a special case of solutions
in \cite{dMFM}). We will study the BLG model 
for this case in \S\ref{sec:2A}.  It demonstrates
the essential feature that the supersymmetric
system acquires mass proportional to eigenvalues of $J$.
However, since we put $f^{ijk}_a=0$, there is no interaction.
In order to have the interacting system, we need
nonvanishing $f^{ijk}_a$. 

For simplicity, 
let us assume that there is a suitable basis of generators 
such that the solution (\ref{e:f}) is simplified as 
\ba \label{e:f-1}
f^{ijk}_a=\left\{
\begin{array}{ll}
f^{ijk}_a \quad&i,j,k\in I_a,\\
0 & \mbox{otherwise},
\end{array}
\right.
\ea
where the indices are divided into
$n$ disjoint sets $I=I_1\cup \cdots \cup I_n$, 
and $f^{ijk}_a$ is the structure constant 
for a Lie algebra $g_a$.

Starting with (\ref{e:f-1}), we can solve all the constraints 
(\ref{f4'})--(\ref{f23'}) as follows, 
while (\ref{e:f-1}) already solves (\ref{f4'})--(\ref{f4-2}). 

Eq. (\ref{f13/f18}) is trivial if two of the indices $a, b, c$ 
are identical. 
Assuming (\ref{e:f-1}), eq. (\ref{f13/f18}) imposes 
no constraint on $J_{ab}^{ij}$ if $i \in I_a$ or $i \in I_b$.
In general, if $f^{ijk}_c \neq 0$ for $c \neq a$ and $c\neq b$, 
then $J_{ab}^{ij} = 0$ if $i \in I_c$. 
Hence we consider the case 
\ba
J_{ab}^{ij} \neq 0 \qquad \mbox{only if} \qquad i, j\in I_a 
\quad \mbox{or} \quad i, j\in I_b. 
\label{alpha}
\ea
Eq. (\ref{f9'}) is now trivial if all indices $a, b, c$ are different. 
If two of the indices are the same, 
it is equivalent to (\ref{f6/f10}). 

According to (\ref{f6/f10}), $J_{ab}$ is 
a derivation for both Lie algebras $g_a$ and $g_b$. 
A derivation ${\cal D}$ is a map from $g$ to $g$ such that
\ba
{\cal D}([e^i, e^j]) = [{\cal D}(e^i), e^j]+[e^i, {\cal D}(e^j)].
\ea
As a result of (\ref{f6/f10}), 
one can define a derivations ${\cal D}_{ab}$ by 
\ba \label{DJ}
{\cal D}_{ab}(e^i) = J^{ij}_{ab} e^j. 
\ea 

The simplest case is when $J_{ab}$ corresponds to 
an inner automorphism, so 
\ba
\label{Jsol}
J_{ab}^{ij} = \Lambda_{ab}^k f_a^{ij}{}_k - \Lambda_{ba}^k f_b^{ij}{}_k, 
\ea
where $\Lambda_{ab}^k = 0$ unless $k\in I_a$.
(Note that the indices $a, b$ are not summed over in (\ref{Jsol}).)
In this case
${\cal D}_{ab}(\cdot) = [(\Lambda_{ab}^k - \Lambda_{ba}^k) e_k,\,\cdot\;]$.
It will be more interesting 
if ${\cal D}_{ab}$ instead corresponds to an infinitesimal outer automorphism 
(an {\em outer derivation}). 
\footnote{We have to keep
in mind that the existence of such automorphisms is quite nontrivial.
We will come back to this issue below.}

If all indices $a, b, c, d$ are all different, 
(\ref{f11-1}) is trivial due to (\ref{alpha}). 
If $a = d \neq b \neq c$, (\ref{f11-1}) says 
that the Lie bracket $[J_{ab}, J_{ac}]$ 
is an inner automorphism. 
The solution of (\ref{f11-1}) is in general given by
\ba \label{Ksol2}
K_{abc} := K_{abc}^i e^i = 
[{\cal D}_{ac}, {\cal D}_{bc}] +
[{\cal D}_{ba}, {\cal D}_{ca}] +
[{\cal D}_{cb}, {\cal D}_{ab}] +
C_{abc}, 
\ea
where the antisymmetric tensor
$C_{abc} = C^i_{abc}$ is a central element in $g$.
Since all derivations of a Lie algebra is always a Lie aglebra, 
the Lie bracket $[{\cal D}_{ab}, {\cal D}_{cd}]$ 
satisfies the Jacobi identity. 

For $J_{ab}$ given by an inner automorphism (\ref{Jsol}), 
$K^i_{abc}$ can be solved from (\ref{f11-1}) to be
\ba
\label{Ksol}
K^i_{abc} = \Lambda_{ab}^j\Lambda_{ac}^k f_a^{ijk}
+ \Lambda_{bc}^j\Lambda_{ba}^k f_b^{ijk}
+ \Lambda_{ca}^j\Lambda_{cb}^k f_c^{ijk} + C_{abc}^i. 
\ea
(Indices $a, b, c$ are not summed over in this equation.)
The term $\Lambda_{ab}^j\Lambda_{ac}^k f_a^{ijk}$
corresponds to the Lie bracket of the two automorphisms 
generated by $\Lambda_{ab}$ and $\Lambda_{ac}$ on $g_a$. 
However, the case of $J_{ab}$ generating an inner automorphism is 
not interesting because $J_{ab}$ and $K^i_{abc}$ 
can be both set to zero after a change of basis 
(\ref{r2},\ref{r3}),
\ba
e'{}^i &=& e^i - \sum_b \Lambda^i_{ab} v^b \quad \mbox{for} \quad i \in I_a, \\
u'_a &=& u_a - \sum_b \Lambda^i_{ba} e^i .
\ea
Therefore, in the following we will focus on the case 
when $J_{ab}$ is an outer automorphism. 

When all indices $a, b, c, d, e$ are different, 
(\ref{f17'}) can be easily satisfied if 
\ba
\label{beta}
C_{abc}^i = 0 \qquad \mbox{unless} \qquad i \in I_a\cup I_b\cup I_c.
\ea
Together with (\ref{alpha}), this implies that
$K_{abc}^i$ (\ref{Ksol2}) vanishes unless $i \in I_a\cup I_b\cup I_c$.

Due to (\ref{alpha}) and (\ref{beta}), 
eq.\,(\ref{f17'}) is trivial if all indices $a, b, c, d, e$ are different. 
If $e = a$, it is 
\ba
K_{abc}^i J_{ad}^{ij} + K_{acd}^i J_{ab}^{ij} + K_{adb}^i J_{ac}^{ij} = 0.
\ea
One can then check that this follows from (\ref{Ksol2}) 
and the constraint
\ba
\label{DC}
{\cal D}_{ab}(C_{acd}) + {\cal D}_{ac}(C_{adb}) + {\cal D}_{ad}(C_{abc}) = 0
\ea
as a result of the Jacobi identity of the Lie bracket of ${\cal D}_{ab}$'s. 
The same discussion applies to (\ref{f15/f20}), (\ref{f17-1}), 
(\ref{f19'}) 
and (\ref{f23'}). 


Before closing this subsection, let us comment on 
infinitesimal outer automorphisms. 
For finite dimensional Lie algebra, 
we have two examples. 
The first example is when the Lie algebra is Abelian, 
and any nontrivial linear map of the generators is an outer automorphism.
The 2nd example is when the Lie algebra is 
that of matrices composed of upper triangular blocks
\ba
\left(\begin{array}{cc} A & B \\ 0 & C \end{array}\right),
\ea
where $A, B, C$ are $m\times m$, $m\times n$ and $n\times n$ matrices, 
respectively.
An arbitrary scaling of the off-diagonal block $B$ is an outer automorphism.
In both of these examples, the coefficients of $e_i$ in 
the expansion of $X^I$ or $\Psi$ 
do not participate in interactions in the BLG model, 
unless $e^i$ is inert to the outer derivation. 
Hence the appearance of outer derivation in these cases 
is irrelevant to physics. 
A nontrivial example is found when $g$ is an infinite dimensional 
Lie algebra. 
This example is studied in \S \ref{D2D3}.

\subsection{Summary of the 3-algebra solutions}
\label{s:sum}

To summarize the result of our construction of a new 3-algebra, 
the general solution of the fundamental identity for our ansatz 
\ba
{}[u_a, u_b, u_c] &=& K^i_{abc} e_i + L_{abcd} v^d, \label{uuu} \\
{}[u_a, u_b, e^i] &=& J_{ab}^{ij} e_j - K^i_{abc} v^c, \label{uue} \\
{}[u_a, e^i, e^j] &=& J_{ab}^{ij} v^b + f_a^{ijk} e_k, \label{uee} \\
{}[e^i, e^j, e^k] &=& - f_a^{ijk} v^a, \label{eee}
\ea
is given by 
(\ref{e:f-1}), (\ref{DJ}) and (\ref{Ksol2}), 
which are repeated here for the convenience of the reader,
\ba
f^{ijk}_a&=&\left\{
\begin{array}{ll}
f^{ijk}_a \quad&i,j,k\in I_a,\\
0 & \mbox{otherwise},
\end{array}
\right. \\
J^{ij}_{ab} e^j &=& {\cal D}_{ab}(e^i) \qquad 
\mbox{for a derivation} \;\; {\cal D}_{ab},  \\
K_{abc} &:=& K_{abc}^i e^i = 
[{\cal D}_{ac}, {\cal D}_{bc}] +
[{\cal D}_{ba}, {\cal D}_{ca}] +
[{\cal D}_{cb}, {\cal D}_{ab}] +
C_{abc}, 
\ea
where $C_{abc}$ are central elements in $g$ satisfying 
(\ref{beta}) and (\ref{DC})
\ba
&C_{abc}^i = 0 \qquad \mbox{unless} \qquad i \in I_a\cup I_b\cup I_c, \\
&{\cal D}_{ab}(C_{acd}) + {\cal D}_{ac}(C_{adb}) + {\cal D}_{ad}(C_{abc}) = 0.
\ea
The nontrivial part of the metric is given by 
\ba
\langle e^i, e^j\rangle= g^{ij}\,,\quad
\langle u_a, v^b \rangle=\delta_a^b \, , \label{metric}
\ea
where $g^{ij}$ is the Killing form of the Lie algebra $g$. 
Although we have assumed that $g^{ij}$ is 
positive definite in the derivation above, 
it is obvious that the 3-algebra can be directly generalized 
to a generic Killing form which is not necessarily positive definite.

Compared with the 3-algebra discovered in \cite{GNP,GMR,HIM}, 
the 3-algebra constructed above contains more information. 
While $e^i$'s are generators of a Lie algebra 
$g=g_1+\cdots+g_n$, 
$J_{ab}$'s correspond to infinitesimal outer automorphisms (outer derivations), 
and $K_{abc}$ encodes both the commutation relations among $J_{ab}$'s 
and choices of central elements in $g$. 

Based on this analysis, we will analyze the BLG model
for some examples of Lorentzian 3-algebras:
\begin{enumerate}
\item $M=2$, $J^{ij}_{ab}=\epsilon_{ab} J^{ij}$ ($i,j=1, \cdots, n$),
others$=0$ (\S\ref{2A}):
This is the simplest finite dimensional example where
some character of the Lorentzian symmetry is displayed.
Namely the BLG model defines the $\cN=8$ supersymmetric
vector multiplets.
\item   $M=2$, $J^{ij}_{ab}=\epsilon_{ab} J^{ij}$, $f^{ijk}_1\neq 0$,
others$=0$ (\S\ref{fneq0}): This is the simplest nontrivial example which contains
the interaction.  We will present our result by studying the
Yang-Mills system (\ref{eLagYM}) 
where the gauge symmetry is defined by
Lorentzian Lie algebra. This is possible since the 3-algebra can be written
in the form (\ref{eLext}).  In such case, one can skip the discussion of
eliminating  one pair of ghost fields.  It also illuminate the structure
of the Yang-Mills system with Lorentzian Lie algebra.
\item Lie 3-algebra associated with affine Kac-Moody Lie algebra (\S\ref{loopalgebra}):
This is the special case of above example where the Lorentzian Lie
algebra is given by the affine Lie algebra. It illuminates how Kaluza-Klein
mass is generated by the ghost fields.
\item Lie 3-algebra associated with general loop algebras (\S\ref{M2Dp}):
By this generalization we describe 
the compactification on general torus $T^p$
with constant $B$ field flux on it. 
\item Lorentzian 3-algebra with $F^{ijkl}\neq 0$
(\S\ref{sM5}):
We give
a brief explanation how construction of M5-brane \cite{HM, HIM, HIMS}
can be related to the Lorentzian 3-algebra (\ref{M5F},\ref{M5f}--\ref{M5K})
and how the analysis in \cite{HM, HIM, HIMS} can be related to the analysis
in this paper.
\end{enumerate}

\section{BLG model for Lorentzian 3-algebra with $J^{ij}_{ab}\neq 0$}
\label{sec:2A}

In this section, we describe generic features of BLG model 
when $J^{ij}_{ab}\neq 0$.
We will first start with the ``minimal'' choice, 
namely we set other structure constants to zero,
\ba
F^{ijkl}=f^{ijk}_a=K^i_{abc}=0\,.
\ea
We note that
this is the simplest example considered in \cite{dMFM}.
For this simplest choice, we see that BLG model
gives rise to a free $\cN=8$ supersymmetric massive gauge theory
after the Higgs mechanism is used to eliminate 
the negative-norm fields. 
After including other structure constants,
we have an interacting theory. 
The direct analysis of interacting model from BLG model itself 
is somehow complicated and less illuminating, 
hence we will consider its equivalent version, 
the super Yang-Mills theory, in the following.

\subsection{Component expansion}
\label{2A}

The BLG action is defined by (\ref{S}--\ref{LCS}), 
with the indices $A, B =(e^i,u_a,v_a)$.
For simplicity, we first study the special case when
the only nonvanishing part of the structure constant of the 3-algebra is
\ba
f^{u_a u_b i j}=\epsilon_{ab}J_{ij},
\ea
where $a,b=1,2$ and $i,j=1,\cdots, n$.

As usual, we expand the relevant
parts of the fields as
\ba
&X^I = X^I_i e^i + X^I_{a}u_a + {\ul X}^I_{a}v_a, \\
&\Psi = \Psi_i e^i + \Psi_{a}u_a + {\ul \Psi}_{a}v_a, \\
&\sum_{ij}J_{ij}A_{\mu ij} =: A'_{\mu},\\
&A_{\mu i u_a}=-A_{\mu u_a i}=:\frac{1}{2} B_{\mu ia},\\
&A_{\mu u_a u_b}=:\frac12 C_\mu\epsilon_{ab} \,.\  
\ea
In terms of the modes, the covariant derivative (\ref{eq:cov}) becomes
\ba
(D_\mu X^I)_i &=& \partial_\mu X^I_i 
+\epsilon_{ab} J_{ij} B_{\mu ja} X_b^I
+C_\mu J_{ij}X^I_j \,, \nt
(D_\mu X^I)_{u_a}&=&\partial_\mu X^I_a\,,\nt
(D_\mu X^I)_{v_a}&=&\partial_\mu \ul X^I_a+
\epsilon_{ab}(A'_\mu X^I_b +J_{ij} B_{\mu bi}X^I_j),
\ea
and similar expressions for $\Psi$.

The Chern-Simons action (\ref{LCS})
can be rewritten in terms of 
the component gauge fields as
\ba
L_{CS}&=& \epsilon^{\mu\nu\lambda}(
A'_\mu \partial_\nu C_\lambda -\frac{1}{2}
J_{ij} B_{\mu ia} (\partial_\nu B_{\lambda jb}+C_{\nu}
J_{jk}B_{\lambda kb}))\nn\\
&=:&\epsilon^{\mu\nu\lambda}(
A'_\mu \partial_\nu C_\lambda -\frac{1}{2}
J_{ij} B_{\mu ia} \hat D_\nu B_{\lambda jb}).
\ea
The gauge field $A'_\mu$ appears only in the Chern-Simons term.
It does not participate in the dynamics but only imposes
the flatness condition $\partial_{[\nu}C_{\mu]}=0$ 
as the equation of motion.

In the original BLG model, the gauge symmetry transformations are
\ba
\delta X_A^I &=& {\ud {\tilde\Lambda} BA} X_B^I, \nt
\delta \Psi_A &=& {\ud {\tilde\Lambda} BA} \Psi_B, \nt
\delta {\dud {\tilde A} \mu BA} &=& 
 \partial_\mu {\ud {\tilde\Lambda} BA}
-{\ud {\tilde\Lambda} BC}{\dud {\tilde A} \mu CA}
+{\dud {\tilde A} \mu BC}{\ud {\tilde\Lambda} CA}.
\ea
We introduce the components of the gauge parameters as
\ba
\Lambda_{u_a u_b}=:\frac{1}{2}\gamma,\qquad
\Lambda_{iu_a}=:\frac{1}{2}\beta_{ia},\qquad
J_{ij}\Lambda_{ij}=:\alpha\,.
\ea
Then the gauge symmetry transformation in terms of the modes becomes
\ba
\delta \Phi_i&=& \epsilon_{ab}J _{ij}\beta_{jb} \Phi_a-\gamma J_{ij} \Phi_j, \nn\\
\delta \Phi_a &=& 0, \nn\\
\delta\ul \Phi_{a}&=& \alpha\epsilon_{ab}\Phi_b -J_{ij}\epsilon_{ab}\beta_{jb} \Phi_a, \\
\delta A'_\mu&=&\partial_\mu \alpha, \nn\\
\delta B_{\mu ib}&=& \partial_\mu \beta_{ib}+J_{ij}\beta_{jb} C_\mu
-J_{ij}B_{\mu jb}\gamma, \nt
\delta C_\mu &=& \partial_\mu\gamma\,,
\ea
where $\Phi=X^I, \Psi$.
The gauge transformations for the gauge fields
$A'_\mu,C_\mu$ associated with the parameters $\alpha,\gamma$ are
Abelian.


In the original BLG model, the supersymmetry transformations are
\ba
\delta X^I_A &=& i\bar\epsilon \Gamma^I \Psi_A, \nn\\
\delta \Psi_A &=& D_\mu X_A^I \Gamma^\mu \Gamma^I \epsilon
-\frac16 X_B^I X_C^J X_D^K {\ud f {BCD} A} \Gamma^{IJK}\epsilon, \nn\\
\delta {\dud {\tilde A} \mu B A}&=& i\bar\epsilon \Gamma_\mu \Gamma_I
X_C^I \Psi_D {\ud f {CDB}A},
\ea
where ${\ud {\tilde\Lambda} BA}=\Lambda_{CD}{\ud f {CDB}A}$. 
So, in terms of the components, the nontrivial parts of the
supersymmetry transformation (namely, for $\Psi$ and $\tilde A_\mu$) become
\ba
\delta \Psi_i &=& D_\mu X_i^I\Gamma^\mu \Gamma^I\epsilon 
+\frac12 \epsilon_{ab}J_{ij} X_{a}^I X_{b}^J X_j^K \Gamma^{IJK}\epsilon,\nt
\delta \Psi_{a} &=& \partial_\mu X_{a}^I \Gamma^\mu \Gamma^I \epsilon, \nt
\delta \ul\Psi_{a} &=& (D_\mu X)_{v_a}\Gamma^\mu \Gamma^I \epsilon 
+\frac12 \epsilon_{ab}J_{ij} X_i^I X_j^J X_{b}^K \Gamma^{IJK} \epsilon, \nt
\delta A'_\mu&=& i\bar\epsilon\Gamma_\mu \Gamma_I X^I_i \Psi_j J_{ij},\nt
\delta B_{\mu ib}&=& i\bar\epsilon \Gamma_\mu\Gamma_I X^I_{[i}\Psi_{b]},\nt
\delta C_{\mu} &=& i\bar\epsilon \Gamma_\mu \Gamma_I X^I_a \Psi_b \epsilon_{ab}\,.
\ea

By the definition of the BLG model, we obtain an $\cN=8$ SUSY system 
with ghost fields.

\subsection{Elimination of ghosts}

Variation of the Lagrangian by fields $\ul X^I, \ul \Psi$ 
gives
\ba
\partial^2 X^{I}_{a}=0,~~~
\Gamma^\mu\partial_\mu\Psi_{a}=0.
\ea
As already reviewed in \S\ref{s:L-rev},
we solve them by the assignment \cite{HIM}
or the introduction of extra gauge symmetry
\cite{Bandres:2008kj,Gomis:2008be}:
\ba
X^I_{a}= \lambda^I_a,~~~
\Psi_{a} = 0;\quad a=1,2\,.
\ea
It is clear that this choice does not break gauge symmetry nor
supersymmetry, since the transformation of these fields is closed. 

The Lagrangian is simplified considerably after inserting these VEV's:
\ba
L&=&-\frac{1}{2}\sum_i(\hat D_\mu X_i^I+\epsilon_{ab} J_{ij} B_{\mu ja}
\lambda^I_b)^2 
+\frac{i}{2}\bar\Psi_i \Gamma^\mu \hat D_\mu\Psi_i\nn\\
&& -\frac{1}{2} (J^2)_{ij}\Delta^2 X^I_i P_{IJ} X^J_j
-\frac{i}2\Delta\bar\Psi_i \Gamma^\|  J_{ij}\Psi_j\nn\\
&&+\epsilon^{\mu\nu\lambda}(
A'_\mu \partial_\nu C_\lambda -\frac{1}{2}
J_{ij} B_{\mu ia} \hat D_\nu B_{\lambda jb}),
\label{Lag2A}
\ea
where
\ba
\hat D_\mu \Phi_i&:=&\partial_\mu \Phi_i + C_{\mu}J_{ij}\Phi_j,
\\
\Delta^2&:=&|\vec \lambda_1|^2 |\vec \lambda_2|^2 
-(\vec \lambda_1\cdot \vec \lambda_2)^2, \\
P_{IJ}&:=&\delta_{IJ}-\sum_{a=1,2} \lambda_a^I \pi_a^J, \\
\vec \pi_1&:=& \frac{1}{\Delta^2}(|\vec \lambda_2|^2\vec \lambda_1
-(\vec \lambda_1 \cdot \vec \lambda_2) \vec \lambda_2),
\quad \vec \pi_2 = (1\leftrightarrow 2)\,,
\\
\Gamma^\| &:=&\frac{1}{2\Delta} \Gamma_{IJ}\epsilon_{ab} \lambda^I_a \lambda^J_b,\quad
(\Gamma^\|)^2=1.
\ea
The $\vec \pi_a$ ($a=1,2$) is the dual basis of $\vec \lambda_a$,
namely
$(\vec \pi_a, \vec \lambda_b)=\delta_{ab}$.
The matrix $P_{IJ}$ is a projector with codimension two
which satisfies
$P\vec \lambda_a=0$ ($a=1,2$) and $P^2=P$.
The potential implies that six components of $X^I$ become massive
after putting VEV to $X_a^I$, while the two components in
the plane spanned by $\vec \lambda_{a}$ remain massless.
Actually the latter can be removed by redefinition of
$B_{\mu ja}$,
\footnote{
If the matrix $J_{ij}$ is not invertible, 
one can first decompose the linear space $\{e^i\}$ into two parts:
the part on which $J_{ij}$ is trivial and the part on which 
$J$ is invertible.
We focus our attention on the latter part.
}
\ba
B'_{\mu ia}&=&B_{\mu ia}+\delta B_{\mu ia}\\
\delta B_{\mu ia}&=& \hat D_\mu \beta_{ia},
\quad
\beta_{ia}:=(J^{-1})_{ij}\epsilon_{ab} \pi_{bJ}X^J_j.
\ea
Since this redefinition takes the form of the gauge transformation
for $B_{\mu ia}$, it does not change the form of Chern-Simons term.
The gauge symmetry associated with $\beta_{ia}$ is fixed by this
manipulation and will not survive in the gauge fixed Lagrangian.

After this gauge transformation, the Chern-Simons Lagrangian
$L_{CS}$ remains the same while the kinetic term for $X$ becomes
\ba
L_X&=&-\frac12\sum_i (\hat D_\mu X^I)_i P_{IJ} (\hat D_\mu X^J)_i 
+\frac{1}{2}\sum B_{\mu ja} (J^2)_{jk} Q_{ab} B^\mu_{kb}, \\
Q_{ab}&:=& \epsilon_{a a'}\epsilon_{b b'} (\vec \lambda_{a'}, \vec \lambda_{b'}).
\ea
The second term in $L_X$ is the mass term for the gauge potential
$B_{\mu ia}$.

To see the mass term for gauge fields
more explicitly, we combine the relevant parts from $L_{CS}$
and $L_X$ to give the action for $B_{\mu ia}$,
\ba
L'_B
= -\frac12 \epsilon^{\mu\nu\lambda} J_{ij} B_{\mu i1}
(F_{\nu\lambda})_{j2} +\frac{1}{2}\sum B_{\mu ja} (J^2)_{jk} Q_{ab} B^\mu_{kb}\,,
\ea
where $(F_{\nu\lambda})_{ja}=(\hat D_\nu B_\lambda)_{ja}-(\hat D_\lambda B_\nu)_{ja}$.
In the second term, we used partial integration.  Since  there are no
derivatives of $B_{\mu i1}$, we integrate over them, and
\ba
L_B\rightarrow \frac{1}{Q_{11}} \left(-
\frac{1}4\sum {(F_{\nu\lambda})_{k2}}^2
+ \frac{1}{2} (J^2)_{ij}\Delta^2 B_{\mu i2}{B^\mu}_{j2}
\right).
\ea

The gauge symmetry is now reduced to Abelian transformations,
\ba
\delta\Phi_i = -\gamma J_{ij} \Phi_j,\quad
\delta B_{\mu ia}= -J_{ij}B_{\mu ja} \gamma,\quad
\delta A'_\mu = \partial_\mu \alpha,\quad
\delta C_\mu = \partial_\mu \gamma\,.
\ea
They are, however, mostly trivial since the gauge field $C_\mu$
which appears in the covariant derivative is required to be flat
by the equation of motion.

In the end, we find that we have $n$ {\em massive} vector fields
$B_{\mu i 2}$, $6n$ massive scalars $P^{IJ} X_i^J=:(X')^I_i$ 
and $8n$ fermion fields $\Psi_i$. 
The mass spectrum of this supersymmetric system 
is given by 
\ba\label{mass0}
m^2=\mbox{eigenvalues of }J^2 \Delta^2.
\ea
We note that this mass formula is invariant under $SL(2,\mathbf{R})$ 
transformations:
\ba
\vec \lambda'_a = g_{ab} \vec \lambda_b,
\quad
g_{ab}\in SL(2,\mathbf{R})\,. 
\ea
This property is natural if we want to associate the system
with $T^2$ compactification of M-theory, 
so that the mass spectrum corresponds to the Kaluza-Klein modes.
This feature becomes
more explicit in the example considered in the next section.

The original supersymmetry remains the same
($\cN=8$) after the Higgs mechanism,
\ba
\delta {X'}_i^I &=& i\bar\epsilon P_{IJ}\Gamma^J \Psi_i, \\
\delta \Psi_i &=& D_\mu X_i^I P_{IJ} \Gamma^\mu \Gamma^J\epsilon 
+\Delta J_{ij}  X_j^I P_{IJ} \Gamma^{J}\Gamma^\|\epsilon, \\
\delta A'_\mu&=& i\bar\epsilon\Gamma_\mu \Gamma^I 
P_{IJ} X^J_i \Psi_j J_{ij}, \\
\delta C_{\mu} &=& 0\,.
\ea

\subsection{Inclusion of $f^{ijk}_a\neq 0$}
\label{fneq0}

By turning on $f^{ijk}_a\neq 0$, 
one may include interacting non-Abelian gauge symmetry in the action.
For simplicity, we set
\ba
f^{ijk}_1\neq 0, \quad
f^{ijk}_2=0, \quad
J^{ij}\neq 0.
\ea
In this case, we can rewrite it as
\ba\label{eLext}
{[u_1,T^A,T^B]}&=&{\ud f {AB}C}T^C, \nt
{[v_1,T^A,T^B]}&=&0, \nt
{[T^A,T^B,T^C]}&=&-h^{CD}{\ud f{AB}D}v_1,
\ea
where $A,B,\cdots=\{u_2,v_2,i\}$, $f^{ijk}:=f^{ijk}_1$ and $f^{u_2ij}:=J^{ij}$.
This algebra is similar to that of \cite{HIM,GMR,Benvenuti:2008bt}, 
that is, a $(u_1,v_1)$-extension of Lie 3-algebra (\ref{L1}). 
A different point is that this Lie 3-algebra $\{T^A\}=\{T^i,u_2,v_2\}$ 
has Lorentzian generators, while that of \cite{HIM,GMR,Benvenuti:2008bt} 
is a standard (positive-definite) Lie algebra.

In this subsection, we denote generators of this algebra as $\{e^i,u,v\}$, instead of $\{T^i,u_2,v_2\}$. Then the metric (or Killing form) and structure constant is
\ba &&
\langle e^i, e^j\rangle=\delta^{ij},\quad
\langle u, v\rangle =1;\nt&&
f^{ijk}, \quad f^{uij}=J^{ij},\quad \mbox{otherwise}=0,
\ea
where $i=1,\cdots, N$.
The Jacobi identity is written as
\ba
&& f^{ijl}f^{lkm}+f^{jkl}f^{lim}+f^{kil}f^{ljm}=0,\\
&& f^{ijl}J^{lk}+f^{jkl}J^{li}+f^{kil}J^{lj}=0,
\ea
which are consistent with the fundamental identity for the Lie 3-algebra $\{T^i,u_{1,2},v_{1,2}\}$.
This is the simplest ``Lorentzian extension" of Lie algebra,
\ba
[e^i, e^j]={f^{ij}}_k e^j+J^{ij} v,\quad
[u, e^i]=J^{ij} e^j.
\ea
This extension is trivial if $J^{ij}$ is an inner automorphism
\ba
J^{ij}={f^{ij}}_k \alpha^k
\ea
for some parameter $\alpha^k$.  One may then redefine the basis
\ba
e'^i=e^i+\alpha^i v,\quad
u'=u-\alpha_i e^i,\quad
v' = v,
\ea
such that the algebra becomes the direct sum of the original Lie algebra
and Lorentzian pairs:
\ba
&& [e'^i, e'^j]={f^{ij}}_k e'^k,\quad
\mbox{other commutators}=0\,;\\
&& \langle e'^i, e'^j\rangle=\delta^{ij},\quad
\langle u', v'\rangle=1,\quad
\mbox{other inner products}=0\,.
\ea
In the following, we will focus on the nontrivial case
where $J$ gives an infinitesimal outer automorphism.

As we explained in \S\ref{s:L-rev}, (according to \cite{HIM},) 
BLG model with Lorentzian Lie 3-algebra results in 
super Yang-Mills theory with Lie algebra. 
So, let us consider the Yang-Mills theory coupled with
scalar fields $X^I$ ($I=1,\cdots n$) and 
spinor fields $\Psi$ based on this extended algebra:
\ba
L&=&
-\frac{1}{2}\langle D_\mu X^I, D^\mu X^I\rangle
+\frac{\lambda_1^2}4 \langle [X^I, X^J],[X^I, X^J]\rangle \nt
&&
+\frac{i}2 \langle \bar\Psi,\Gamma^\mu D_\mu \Psi \rangle
+\frac{i\lambda_1}2 \langle \bar\Psi,\Gamma_I [X^I, \Psi] \rangle
-\frac{1}{4\lambda_1^2}\langle F_{\mu\nu} F^{\mu\nu}\rangle\\
&=:&L_X+L_{pot}+L_\Psi+L_{int}+L_A\,,
\ea
where $X^I$ takes the adjoint representation
\ba
X^I &=& X^I_i e^i +X^I_u u+X^I_v v,
 \\
(D_\mu X^I)_i &=& \partial_\mu X^I_i - {f^{jk}}_i A_{\mu j}X^I_k
- J^{ji} C_\mu X_j^I+J^{ji} A_{\mu j} X^I_u \nn\\
&=:& (\hat D_\mu X^I)_i +J^{ji} A_{\mu j} X^I_u, \\
(D_\mu X^I)_u&=& \partial_\mu X^I_u, \\
(D_\mu X^I)_v&=& \partial_\mu X^I_v+J^{ij} A_{\mu i} X^I_{j}\,,\\
A_{\mu u}&=:& C_\mu,\quad
A_{\mu v} =: B_\mu
\ea
and similar expressions for $\Psi$.
The covariant derivative corresponding to the gauge symmetry 
generated by $e^i$ should thus be defined as
\ba
\hat{D}_{\mu} = \partial_{\mu} - C_{\mu}{\cal D}_u - A_{\mu i} e^i, 
\label{hatD}
\ea
where ${\cal D}_u$ is the derivation defined by $J$:
\ba
{\cal D}_u(e^i) = J^{ij} e^j. 
\ea
On the right hand side of (\ref{hatD}), $e^i$ is used to imply
the adjoint action of $e^i$, namely $e^i(x)=[e^i, x]$.
The gauge transformation is written as
\ba
\delta \Phi_i &=& {f^{jk}}_i \epsilon_j \Phi_k+J^{ki}\gamma \Phi_k
-J^{ji} \epsilon_j \Phi_u, \\
\delta \Phi_u &=& 0, \\
\delta A_{\mu i}&=& \partial_\mu\epsilon_i +{f^{jk}}_i \epsilon_j
A_{\mu k}+ J^{ki}\gamma A_{\mu k}-J^{ji}\epsilon_j C_\mu \nt
&=:& (\hat D_\mu \epsilon)_j +J^{ji}\gamma A_{\mu j}
\ea
for $\Phi=X^I,\Psi$.

The kinetic term for $X^I$ becomes
\ba
L_X=\frac12 (\hat D_\mu X^I_i + J^{ji} A_{\mu j} X^I_0)^2
+ \partial^\mu X^I_u (\partial_\mu X^I_v-J^{ij} A_{\mu i} X^I_{j})\,.
\ea
The variation of $X^I_v$ gives $\partial^2 X^I_u=0$.  So we take
it as constant as before,
\ba
X^I_u=\lambda_2 \delta_{I1}\,.
\ea
After imposing this VEV, 
\ba
L_X=-\frac12\sum_{I'=2}^n  (\hat D_\mu X^{I'}_i)^2
-\frac1{2\lambda_1^2} F_{\mu u}^2, 
\label{lx}
\ea
where 
\ba
F_{\mu u} &:=& [\hat{D}_{\mu}, \hat{D}_u], \\
\hat{D}_u &:=&\lambda_1( \lambda_2 {\cal D}_u + X^1_i e^i ).
\ea
We are thus led to interpret ${\cal D}_u$ (or $J$) 
as the derivative of a certain noncommutative space 
in the direction of $X_u$.
The situation here is reminiscent of the result of quotient conditions 
in the context of Matrix Models in dealing with 
orbifolds and orientifolds \cite{HWW}.
In analogy, 
since we have taken the VEV of $X_u$ to be in the direction of $X^1$, 
$X^1_j$ plays the role of a gauge potential 
and $J_{ij}$ that of a covariant derivative 
on a noncommutative space, and thus 
$\hat{D}_u$ mimics a covariant derivative.
We will see in the next section that for the compactification on a circle,
$\hat{D}_u$ is indeed the covariant derivative in the compactified direction.

If we fix the gauge by $X^1_i = 0$, 
the second term in (\ref{lx}) becomes
\ba\label{mass}
-\frac{\lambda_2^2}2(J^2)_{ij} A_{\mu i}A_{\mu j}\,.
\ea
This is the mass term for vector bosons.

The potential term is
\ba
L_{pot} = \frac{\lambda_1^2}4 \sum_{I', J'=2}^n [X^{I'}, X^{J'}]^2
-\frac{1}2 \sum_{J'=2}^n 
(\hat{D}_u X^{J'})^2 .
\ea
If we gauge away $X^1_i$ using the gauge symmetry, 
the last term above is simply 
\ba
- \frac{\lambda_1^2 \lambda_2^2}2 \sum_{J'=2}^n (J^2)_{ij} X^{J'}_i X^{J'}_j .
\ea
It gives the mass term for $X^{J'}$ with exactly 
the same mass as eq.(\ref{mass0}) with $\Delta=\lambda_1\lambda_2$. 
\footnote{
If $J$ is an inner automorphism, i.e. $J^{ki}={f^{jk}}_i \mu_j$,
one may shift $X^1_j= -\mu_j$ to absorb $J$ in $X^1$.
This is consistent with our comment above that 
$J$ can be redefined away if it corresponds to an inner automorphism.
}

The kinetic term for the gauge field becomes
\ba
-\frac{1}{4\lambda_1^2}\langle F_{\mu\nu}, F^{\mu\nu}\rangle
= -\frac{1}{4\lambda_1^2}\left\{
(F_{\mu\nu i})^2 +F_{\mu\nu u} {F^{\mu\nu}}_v
\right\} ,
\ea
where
\ba
F_{\mu\nu i}&=& \partial_\mu A_{\nu i} -\partial_\nu A_{\mu i}
-{f^{jk}}_i A_{\mu j} A_{\nu k} + J^{ij}(C_\mu A_{j\nu} -C_{\nu} A_{j\mu}), \\
F_{\mu\nu u}&=& \partial_\mu C_{\nu} -\partial_\nu C_{\mu}, \\
F_{\mu\nu v}&=& \partial_\mu B_{\nu} -\partial_\nu B_{\mu}
-J^{ij} A_{\mu i} A_{\nu j}\,.
\ea
Variation of gauge field $B_\mu$ gives a free equation of motion for 
$C_\mu$,
\ba\label{eqC}
\partial^\mu \partial_{[\mu} C_{\nu]}=0\,.
\ea
If we start from the BLG action (\ref{Lag2A}), 
we have slightly different Lagrangian,
\ba
L_{A'C}= \epsilon_{\mu\nu\lambda} A'_\mu \partial_\nu C_\lambda ,
\ea
where $A'_\mu$ is an auxiliary field.  
From the viewpoint of the SYM, 
although it is not present from the beginning, 
one can add this term as a way to gauge the global symmetry 
of translation of $C_{\mu}$, 
analogous to (\ref{Xugauge}), 
where we gauged the translation of $X_u$ and $\Psi_u$.
By variation of $A'_\mu$,
$C_\mu$ becomes topological and pure gauge. 
Hence we should set $C_{\nu}$ to be a constant.
It can be interpreted as the projection of the ``$u$''-direction 
on the D-brane worldvolume, 
while $X^I_u$ is the projection of the $u$-direction
in the transverse directions.

On the fermionic parts, after setting the VEV to $\Psi_u = 0$, they become
\ba
L_\Psi=\frac{i}2 \langle \bar\Psi,\Gamma^\mu \hat D_\mu \Psi \rangle,
\ea
and
\ba
L_{int}=\sum_{I'=2}^n 
\frac{i\lambda_1}2\langle\bar\Psi_i,\Gamma_{I'}[X_j^{I'},\Psi_k]\rangle
+\frac{i}2\bar\Psi_i\Gamma_1\hat{D}_u \Psi_i .
\ea
In the gauge $X^1_i = 0$, 
the second term becomes the mass term for 
the fermions with their masses given by the matrix $\lambda_1 \lambda_2 J$. 

To summarize, in the gauge $X^1 = 0$, 
\ba
L&=& L_X+L_\Psi+L_{int}+L_A, \\
L_X&=&\sum_{I',J'=2}^n -\frac12(\hat D
_\mu X_i^{I'})^2
+\frac{\lambda_1^2\lambda_2^2}2 X_i^{I'}(J^2)_{ij}X_j^{I'}, \\
L_\Psi&=&\sum_{I'=2}^n \frac{i}2\bar\Psi
\Gamma^\mu \hat D_\mu
\Psi-\frac{\lambda_1\lambda_2}{2}\bar\Psi_i(i\Gamma_1)J^{ij}\Psi_j , \\
L_{int}&=&\sum_{I', J'=2}^n \frac{\lambda_1^2}4 [X^{I'},X^{J'}]^2
+\frac{i\lambda_1}2 \langle \bar\Psi,\Gamma_{I'}[X^{I'},\Psi] \rangle, \\
L_A &=& -\frac{1}{4\lambda_1^2}F_{\mu\nu}^2-\frac{\lambda_2^2}{2}(J^2)_{ij}A'_{\mu i}A'_{\mu j},
\ea
which is of the form of a massive super Yang-Mills theory 
with the mass matrix $\lambda_1\lambda_2 J_{ij}$.

\section{Application to toroidal compactification of M/string theories}
\label{D2D3}

In this section we first consider an example of the general theory 
studied in \S \ref{fneq0}.
We consider the Kac-Moody algebra as an example of 
the Lorentzian extension of a Lie algebra, 
and show in \S \ref{loopalgebra}
that the SYM theory with the gauge symmetry 
generated by the Kac-Moody algebra is equivalent to 
a SYM theory with a finite dimensional gauge group 
on a base space of higher dimensions. 
Finally, to be complete, in \S \ref{M2Dp}
we consider the BLG model with the full 3-algebraic structure 
to describe M2-branes in flat spacetime 
compactified on a $d$-dimensional (noncommutative) torus 
with background fields.

\subsection{D$p$ to D$(p+1)$ via Kac-Moody algebra}
\label{loopalgebra}

Before we go to the general discussion, let us briefly
consider a simple case where Lie 3-algebra is defined as
(\ref{L1}) where Lie algebra $\mathcal{G}$ itself is a Lorentzian
Lie algebra.  The simplest example is when $\mathcal{G}$
is the affine Lie algebra $\hat g$,
\ba
&& [u, T^a_m]=m T^a_m, \\
&& [T^a_m, T^b_n]= mv g^{ab}\delta_{m+n} +i{f^{ab}}_c T^c_{m+n}, \\
&& [v, u]=[v, T^a_m]=0\, ,
\ea
where $a,b,c=1,\cdots, \mbox{dim}(g)$, $n,m\in\mathbf{Z}$
and $g^{ab}$ is the Killing form of a compact Lie algebra $g$.
This algebra has an invariant metric
\ba
\langle T^a_m, T^b_n\rangle = g^{ab}\delta_{m+n},\quad
\langle u, v\rangle =1\,.
\ea
We note that the generator $v$ is the center of Kac-Moody algebra
and usually taken as a quantized c-number.  Here we identify it
as a nontrivial generator. 
On the other hand, the generator $u$ gives the level (or $-L_0$ in the Virasoro algebra). 
While $T^a_n$ has a positive-definite
metric, the generators $u, v$ have a negative-norm generator. 
\footnote{We note that a different type of Lie 3-algebra based on 
Kac-Moody symmetry was obtained in \cite{Lin:2008qp}.}

We follow the method in \S\ref{fneq0} where we use the super Yang-Mills system on D2 with gauge symmetry $\hat g$
by using the Higgs mechanism for one Lorentzian pair.

In fact, the following analysis can be carried out
for any D$p$-brane system and provides a general mechanism
of the gauge theory with affine gauge symmetry.
What we are going to show is that
the D$p$-brane system whose gauge symmetry is $\hat g$
can be identified with D$(p+1)$-brane system with Lie algebra $g$.

If we start from the BLG model directly, we have a different perspective
in which we will treat more general argument given in the next subsection.

We start from the action
\ba\label{oLag}
L&=&
-\frac{1}{4\lambda^2} \langle F_{\mu\nu} , F^{\mu\nu}\rangle
-\frac{1}{2}\langle D_\mu X^I, D^\mu X^I\rangle
+\frac{\lambda^2}{4}\langle [X^I, X^J],[X^I, X^J]\rangle
\nt&&
+\frac{i}{2} \bar\Psi\Gamma^\mu D_\mu \Psi
+\frac{i\lambda}{2}\bar\Psi \Gamma_I[X^I, \Psi] \,,
\ea
where $X^I(x)$ ($I=1,\cdots, D$) are the scalar field and $\Psi(x)$ is the spinor field. 
Both are in the adjoint representation of $g$.
The world volume index is given as  $\mu,\nu=0,\cdots, p$.
The covariant derivative and the field strength are defined (only in this subsection) as
\ba
D_\mu \Phi:=\partial_\mu \Phi-i[A_\mu,\Phi]\,,\quad
F_{\mu\nu}:=\partial_\mu A_\nu - \partial_\nu A_\mu -i[A_\mu,A_\nu]\,
\ea
for $\Phi=X^I,\Psi$.
The convention here differs from that in \S\ref{s:L-rev}; 
here $A_\mu$ is Hermitian.
We consider the following component expansion,
\ba
 A_\mu &=& A_{\mu(a,n)}T^a_n +B_\mu v+C_\mu u, \\
 X^I &=& X^I_{(a,n)} T^a_n + X^I_u u+ X^I_v v, \\
 \Psi&=&\Psi_{(a,n)}  T^a_n + \Psi_u u+ \Psi_v v\,.
\ea
Various components of the covariant derivative and the field strength are given as
\ba
(D_\mu X^I)_{(an)} &=& \partial_\mu X^I_{an} +{f^{bc}}_a \sum_m A_{\mu (b,m)} X^I_{(c,n-m)}
-n C_\mu X^I_{(a,n)} \nt&& +in A_{\mu (a,n)} X^I_u\nn\\
&=:& (\hat D_\mu X^I)_{(a,n)} + inA_{\mu (a,n)} X^I_u, \\
(D_\mu X^I)_u &=& \partial_\mu X^I_u , \\
(D_\mu X^I)_v &=& \partial_\mu X^I_v + 
\sum_m i m g^{ab} A_{\mu (a, m)} X^I_{(b,-m)}, \\
(F_{\mu\nu})_{(a,n)} &=& \partial_\mu A_{\nu (a,n)}-\partial_\nu A_{\mu (a,n)}
+{f^{bc}}_a \sum_m A_{\mu (b,m)} A_{\nu (c,n-m)}, \\
(F_{\mu \nu})_u &=& \partial_\mu C_\nu -\partial_\nu C_\mu, \\
(F_{\mu\nu})_v &=& \partial_\mu B_\nu- \partial_\nu B_\mu 
+ \sum_m i m g^{ab} A_{\mu (a,m)} A_{\nu (b,-m)}\,,
\ea
and similar expressions for $D_\mu\Psi$.
{}From the kinetic part for $u,v$ components, the equations of motion
for $X_u$, $\Psi_u$ and $C_\mu$ are free,
\ba \label{eq:XC}
\partial^\mu \partial_\mu X_u^I
=\Gamma^\mu \partial_\mu \Psi_u
=\partial^\mu(\partial_\mu C_\nu-\partial_\nu C_\mu)
=0\,.
\ea
We fix their values as
\ba
X^I_u=\mbox{const.}=: \lambda' \delta^{ID},\quad
\Psi_u=0,\quad
\partial_\mu C_\nu-\partial_\nu C_\mu=0\,.
\ea
For the first two relations, we need to use the method
\cite{Bandres:2008kj,Gomis:2008be} as reviewed in \S\ref{s:L-rev}.
We need to introduce the extra gauge symmetry
as commented in the paragraph after (\ref{eqC})
to derive the last one.
For general world volume dimensions, the additional action is
\ba
S_{additional}=-\frac{1}{4\lambda^2} D_{\mu\nu}
(\partial_\mu C_\nu-\partial_\nu C_\mu)\,,
\ea
where $D_{\mu\nu}$ is a new field.  It gives rise to a new gauge symmetry,
\ba
\delta D_{\mu\nu}=\partial_\mu \Xi_\nu -\partial_\nu \Xi_\mu,\quad
\delta B_{\mu}=-\Xi_\mu
\ea
by which we can gauge fix $B_{\mu}=0$.  The equation of motion
by the variation of $D_{\mu\nu}$ gives  the flatness condition
of $C_{\mu}$.

Since the gauge field $C_\mu$ is essentially flat, 
we can ignore it for simplicity
(namely set $C_\mu=0$).  After this, the ghost fields
$C_\mu, B_\mu, X^I_u, X^I_v, \Psi_u, \Psi_v$ disappear from the action, and the system is unitary.

We identify the infinite components of the scalar, spinor and gauge fields as fields in $p+2$ dimensions,
\ba
&&\tilde X^I_{a}(x,y) = \sum_m X^I_{(a, n)}(x) e^{-iny/R}\,,\quad
\tilde \Psi_{a}(x,y) = \sum_m \Psi_{(a, n)}(x) e^{-iny/R}\,,\nt
&&\tilde A_{\mu a}(x,y) =\sum_m A_{\mu (a, n)} (x) e^{-iny/R}, 
\ea
where an extra coordinate $y$ is introduced to parametrize 
$S^1$ with the radius $R$.
We also rename
\ba
\label{XA}
\tilde X^D_a(x,y)\rightarrow  \frac{1}{\lambda} \tilde A_{ya}(x,y)\,.
\ea
The kinetic term of the scalar field $X^I$ can be rewritten as
\ba
-\frac{1}{2} \int \frac{dy}{2\pi R} \left[\sum_{I=1}^{D-1}
(\partial _\mu \tilde X^I_a-{f^{bc}}_a \tilde A_{\mu b} \tilde X^I_c)^2
+ \frac{1}{\lambda^2} \tilde F_{\mu ya}^2
\right], 
\label{Fmuy}
\ea
where
\ba
\tilde F_{\mu ya}:= \partial_\mu \tilde A_{y a}-\partial_y \tilde A_{\mu a}
+{f^{bc}}_a \tilde A_{\mu b}\tilde A_{y c}\,.
\ea
Here the second term can be produced properly if we identify
\ba\label{rKK}
R=1/\lambda\lambda'\,.
\ea
This relation seems strange if we compare with (\ref{lamR}).
It can be fixed by applying the T-duality 
transformation \cite{Taylor:1996ik}.

The second term in (\ref{Fmuy}), 
when combined with the kinetic term for gauge fields,
properly reproduces the kinetic term for $p+2$ dimensional world volume.
The Kaluza-Klein mass from the compactification radius 
(\ref{rKK}) is $n\lambda\lambda'$ which is consistent with the result 
(\ref{mass0}).

Similarly, we can rewrite the commutator term,
\ba
\frac{\lambda^2}{4}\sum_{I,J=1}^D \langle [X^I, X^J],[X^I, X^J]\rangle
&=& \frac{\lambda^2}{4}\sum_{I,J=1}^{D-1} 
\int\frac{dy}{2\pi R} 
\langle [\tilde X^I, \tilde X^J],[\tilde X^I, \tilde X^J]\rangle
\nt &&
-\frac{1}{2} \sum_{I=1}^{D-1}\int \frac{dy}{2\pi R}
(D_y \tilde X^I)^2 .
\ea
Here again the second term can be combined with the kinetic term
for $X^I$ to give the kinetic energy on $p+2$ dimensional world volume.

Finally, we can rewrite the interaction term,
\ba \label{eq:KM-int}
\frac{i\lambda}{2}\sum_{I=1}^D \bar\Psi\Gamma_I[X^I,\Psi]
=
\frac{i\lambda}{2}\sum_{I=1}^{D-1}\int\frac{dy}{2\pi R} 
\bar{\tilde\Psi}\Gamma_I[\tilde X^I,\tilde\Psi]
+\frac{i}{2}\int\frac{dy}{2\pi R} 
\bar{\tilde\Psi}\Gamma^y D_y \tilde \Psi.
\ea
Here, this time, the second term can be combined with 
the kinetic term for $\Psi$.
\footnote{We should notice the definition of 
$\Gamma_\mu$ and $\Gamma_I$ here. 
We see from the kinetic term of $\Psi$ in the Lagrangian (\ref{oLag}) 
that $\Gamma_\mu$ satisfies
$\{\Gamma_\mu,\Gamma_\nu\}=\mbox{diag.}\,(+-\cdots-).$
On the other hand, $\Gamma_I$ should satisfy
$\{\Gamma_I,\Gamma_J\}=\delta_{IJ}$ as usual.
So we choose $\Gamma^D=-i\Gamma^y$ 
and obtain (\ref{eq:KM-int}).}
In the end, the Lagrangian thus obtained is the same
as the original Lagrangian (\ref{oLag}) except that
we change the dimension parameter $D\rightarrow D-1$ and
$p\rightarrow p+1$ and the gauge symmetry 
$\mathcal{G}=\hat g \rightarrow g$:
\ba
L&=&L_A+L_X+L_\Psi+L_{pot}+L_{int} ,\\
L_A&=&-\frac1{4\lambda^2}\int\frac{dy}{2\pi R}
(\tilde F_{\mu\nu}^2+2\tilde F_{\mu y}^2),\\
L_X&=&-\frac12\int\frac{dy}{2\pi R}\sum_{I=1}^{D-1}\left[
 (D_\mu \tilde X^I)^2+(D_y \tilde X^I)^2 \right] ,\\
L_\Psi&=&\frac{i}{2}\int\frac{dy}{2\pi R}
\bar{\tilde\Psi}(\Gamma^\mu D_\mu+\Gamma^y D_y)\tilde\Psi ,\\
L_{pot}&=&\frac{\lambda^2}{4}\sum_{I,J=1}^{D-1} 
\int\frac{dy}{2\pi R} 
\langle [\tilde X^I, \tilde X^J],[\tilde X^I, \tilde X^J]\rangle ,\\
L_{int}&=&\frac{i\lambda}{2}\sum_{I=1}^{D-1}\int\frac{dy}{2\pi R} 
\bar{\tilde\Psi}\Gamma_I[\tilde X^I,\tilde\Psi].
\ea

\subsection{M2 to D$p$ via 3-algebra}
\label{M2Dp}

Here we consider essentially the same physical system 
as the previous subsection, namely 
the compactification of D2-branes on torus, 
but we start from the BLG model for multiple M2-branes 
corresponding to an example of the Lie 3-algebra 
summarized in (\ref{uuu})--(\ref{metric}). 
The formulation here will be more general than above
as we will turn on noncommutativity 
and a gauge field background.

We start by defining a Lie algebra $g_0$
with generators $T^i_{\vec{m}}$, structure constants 
\ba
f^{(i\vec{l})(j\vec{m})(k\vec{n})} = f^{ijk}_{\vec{l}\vec{m}}
\delta^{\vec{l}+\vec{m}+\vec{n}}_{\vec{0}}, 
\label{f}
\ea
and metric
\ba
g^{(i\vec{m})(j\vec{n})} = g^{ij}_{\vec{m}} \delta^{\vec{m}+\vec{n}}_{\vec{0}}.
\ea
Here $\vec{m}$ is a $d$-dimensional vector of integers. 

The simplest example of $g_0$ has
\ba \label{Te}
T^{i}_{\vec{m}} = T^i e^{i\vec{m}\cdot\vec{x}}, 
\ea
where $T^i$ is the generator for $U(N)$ 
and $\vec{x}$ is the coordinate on a $d$-dimensional torus. 
More generally, one can consider a twisted bundle 
on a noncommutative torus $T^d_{\theta}$. 
In this case
\ba \label{TZ}
T^i_{\vec{m}} = T^i Z_1^{m_1}\cdots Z_d^{m_d}, 
\ea
where $T^i$ denotes a generator of the $U(N)$ gauge group, 
and $Z_i$ are noncommutative algebraic elements satisfying
\ba
Z_i Z_j = e^{i\theta'_{ij}} Z_j Z_i.
\ea
The parameter $\theta'$ is in general not the same as 
the noncommutative parameter $\theta$
of the noncommutative torus $T^d_\theta$, 
and it depends on the rank of the gauge group and its twisting. 
$Z_i$ maps a section of the twisted bundle to another section. 
For the trivial bundle, $Z_i = e^{ix_i}$ and (\ref{TZ}) reduces to (\ref{Te}).
The case of $d = 2$ was studied in \cite{Ho:1998hqa,Brace:1998ku}. 
It is straightforward to generalize it to arbitrary dimensions. 

Since the structure constant (\ref{f}) of $g_0$ has the property 
\ba
f^{(i\vec{l})(j\vec{m})(k\vec{n})} \propto
\delta^{\vec{l}+\vec{m}+\vec{n}}_{\vec{0}}. 
\ea 
$g_0$ has derivations 
\ba
\label{J0a}
J_{0a}^{(i\vec{m})(j\vec{n})} = m_a \delta^{(i\vec{m})(j\vec{n})}.
\ea

Now we consider the 3-algebra with the underlying Lie algebra 
$g=g_0$ and $I_{a\neq 0}$'s empty. 
We take $J_{ab} = 0$ if $a, b \neq 0$, 
and $J_{0a}$ given by (\ref{J0a}).
It follows that the first 3 terms in (\ref{Ksol2}) vanish,
hence 
\ba
K^{(i\vec{m})}_{abc} = \delta^i_0\delta^{\vec{m}}_{\vec{0}} C_{abc}, 
\ea
assuming that $T^0$ is the identity of $U(N)$, 
so that $T^{(0\vec{0})}$ is the identity of $g_0$.
In the following we choose 
\ba
K_{0ab}^{i\vec{m}} &=& \delta^i_0 \delta^{\vec{m}}_{\vec{0}} C_{ab}, \\
K_{abc}^{i\vec{m}} &=& 0, \qquad \mbox{otherwise}.
\ea
It will be shown below that the constants $C_{ab}$ 
corresponds to a nontrivial gauge field background. 

The 3-algebra is defined by the 3-brackets
\ba
{}[u_0, u_a, u_b] &=& C_{ab} T^0_{\vec{0}}  
+ L_{0abc} v^c, \label{u0uaub} \\
{}[u_0, u_a, T^i_{\vec{m}}] &=&  m_a T^i_{\vec{m}}
- \delta^i_0 \delta^{\vec{0}}_{\vec{m}} C_{ab} v^b, \\
{}[u_0, T^i_{\vec{m}}, T^j_{\vec{n}}] &=&
m_a g^{ij}_{\vec{m}}\delta^{\vec{0}}_{\vec{m}+\vec{n}} v^a 
+ f^{ijk}_{\vec{m}\vec{n}} T^k_{\vec{m}+\vec{n}}, \\
{}[T^i_{\vec{l}}, T^j_{\vec{m}}, T^k_{\vec{n}}] &=& 
- f^{ijk}_{\vec{l}\vec{m}}
\delta_{\vec{l}+\vec{m}+\vec{n}}^{\vec{0}} v^0, \label{TTTfdv}
\ea
where $a, b, c = 0, 1, 2, \cdots, d$ and $i, j, k = 1, 2, \cdots, N$. 
(Note that we have changed the range of indices $a, b, c$ 
from the convention used above.)

This 3-algebra is actually precisely the Lorentzian algebra
discovered in \cite{GNP,GMR,HIM} constructed from 
the (multiple) loop algebra defined by 
\ba
{}[u_a, u_b] &=& C_{ab} T^0_{\vec{0}} + L_{0abc} v^c, 
\label{loopalg0} \\
{}[u_a, T^i_{\vec{m}}] &=&  m_a T^i_{\vec{m}}
- K^i_{0ab} v^b, \label{loopalg1}\\
{}[T^i_{\vec{m}}, T^j_{\vec{n}}] &=&
m_a g^{ij}\delta^{\vec{0}}_{\vec{m}+\vec{n}} v^a  
+ f^{ij}_{\vec{m}\vec{n}}{}_k T^k_{\vec{m}+\vec{n}}, 
\label{loopalg2} \\
{}[v^a, T^i_{\vec{m}}] &=& 0, \label{loopalg3}
\ea 
where $(a, b = 1, \cdots, d)$.
In the sense that one can construct the 3-algebra 
(\ref{u0uaub})--(\ref{TTTfdv})
from a Lie algebra by adjoining two elements $(u_0, v^0)$,
this 3-algebra is not a good representative of the new class of 3-algebras 
defined in (\ref{uuu})--(\ref{metric}). 
However, it is still a good example 
because it demonstrates the roles played by 
the new parameters $J_{ab}$ and $K_{abc}$, 
which encode the information about derivatives of the Lie algebra $g$, 
which is a subalgebra of the loop algebra (\ref{loopalg0})--(\ref{loopalg3}). 

It follows from the result of \cite{HIM} that 
the BLG model with the Lie 3-algebra (\ref{u0uaub})--(\ref{TTTfdv})
is exactly equivalent to the SYM theory defined 
with the Lie algebra (\ref{loopalg0})--(\ref{loopalg3}). 
In \S \ref{loopalgebra}, 
we showed explicitly that for $d=1$ 
the resulting SYM theory is 
the low energy theory for D$3$-branes. 
Now we briefly sketch the derivation for generic $d$ 
to obtain the SYM theory for D$(d+2)$-branes. 


Expanding the fields in the BLG model, we have
\ba
X^I &=& \sum_{a=0}^{d} X^I_a u_a
+ \hat{X}^I(Z) + Y^I_a v_a, \\
\Psi &=& \sum_{a=0}^{d} \Psi_a u_a
+ \hat{\Psi}(Z) + \Phi_a v_a, \\
A_{\mu} &=& \frac{1}{2} \sum_{a,b=0}^{d} A_{\mu ab} u_a\wedge u_b
+ \sum_{a=0}^d u_a \wedge \hat{A}_{\mu a}(Z) 
+ \sum_{a = 0}^d v^a \wedge \hat{A}'_{\mu a}(Z) \nn \\
&& + \frac{1}{2} \sum_{a, b=0}^d A'_{\mu ab} v^a \wedge v^b
+ \frac{1}{2} \sum_{ij} A_{\mu (i\vec{m})(j\vec{n})} 
T^i_{\vec{m}} \wedge T^j_{\vec{n}} ,
\ea
where we have used (\ref{TZ}) and the notation
\ba
\hat{X}^I(Z) &:=& 
\sum_{\vec{m}} X^{I}_{(i\vec{m})} T^i Z^{m_1}\cdots Z^{m_d}, \\
\hat{\Psi}(Z) &:=& 
\sum_{\vec{m}} \Psi_{(i\vec{m})} T^i Z^{m_1}\cdots Z^{m_d}, \\
\hat{A}_{\mu a}(Z) &:=& 
\sum_{\vec{m}} A_{\mu a(i\vec{m})} T^i Z^{m_1}\cdots Z^{m_d}, \\
\hat{A}'_{\mu a}(Z) &:=& 
\sum_{\vec{m}} A'_{\mu a(i\vec{m})} T^i Z^{m_1}\cdots Z^{m_d}, 
\ea
and $X^I_i(Z)$, $\Psi_i(Z)$ $\hat{A}_{\mu a}(Z)$ and $\hat{A}'_{\mu a}(Z)$ 
are sections of a twisted bundle on $T^d_{\theta}$. 

As we have done it many times already, 
we fix the coefficients of $u_a$ as 
\ba 
X^I_a = \mbox{constant}, \qquad 
\Psi_a = 0, \qquad 
A_{\mu ab} = 0, \qquad (a, b = 0, 1, \cdots, d,)
\ea
and the coefficients of $v_a$ can be ignored. 
Here $A_{\mu ab}$ is chosen to be zero for simplicity. 
If $A_{\mu ab}$'s are nonzero, 
it corresponds to turning on a constant background field strength 
with nonvanishing components of $F_{\mu I}$.

To proceed, we first define covariant derivatives ${\cal D}_a$ 
on the noncommutative torus, such that 
\ba
{}[{\cal D}_a, Z_1^{m_1}\cdots Z_d^{m_d}] 
&=& m_a Z_1^{m_1}\cdots Z_d^{m_d}, \\
{}[{\cal D}_a, {\cal D}_b] &=& C_{ab}, 
\ea
where $C_{ab}$ is the constant background field strength
that determines the twisting of the bundle on $T^d_{\theta}$. 

The rest of the derivation is essentially the same 
as \S \ref{loopalgebra}.
Finally, after integrating out the field $\tilde{A}$, 
the BL Lagrangian turns into that of a SYM theory
\ba
{\cal L} = -\frac{1}{4} \sum_{A, B = 0}^{9} \langle F_{AB}, F^{AB} \rangle 
+ \frac{i}{2} \langle \bar{\Psi}, \Gamma^A \hat{D}_A \Psi \rangle, 
\ea
where 
\ba
F_{\mu\nu} &:=& [\hat{D}_{\mu}, \hat{D}_{\nu}], \\
F_{\mu I} &:=& [\hat{D}_{\mu}, \hat{D}_I], \\
F_{IJ} &:=& [\hat{D}_I, \hat{D}_J] + C_{IJ},
\ea
and
\ba
\hat{D}_{\mu} &:=& \partial_{\mu} - \hat{A}_{\mu 0}(Z), \\
\hat{D}^I &:=& X^I_a {\cal D}_a - \hat{X}^I(Z), \\
C^{IJ} &:=& X^I_a X^J_b C_{ab}. 
\ea
Roughly speaking, only $d$ of the $\hat{D}^I$'s 
are covariant derivatives and the rest $7-d$ are scalar fields. 
To turn on the background field $C_{\mu I}$, 
we can assign nonzero values to $A_{\mu 0a}$ and $A_{\mu ab}$. 

\subsection{M2 to M5 revisited}
\label{sM5}
As we discussed in \S\ref{Fneq0}, there is an
interesting Lorentzian 3-algebra associated with
the Nambu-Poisson bracket on $T^3$ 
defined through the structure constants
(\ref{M5F},\ref{M5f}--\ref{M5K}).
We claim that the BLG model associated with
this 3-algebra is exactly the description
of M5-brane in \cite{HM, HIM, HIMS}
while it was not explicitly understood.
We would like to give a brief sketch on this point.

The key observation to define 6-dimensional fields on M5 from BLG model
is to use the ``mode expansion" such as
\ba
X^I_i(x) T^i \rightarrow X^I_i(x) \chi^i(y)=: X^I(x,y).
\ea
If we add three pairs of Lorentzian generators $(u^a
, v^a)$,
we have to redefine the above expansion as
\ba
\tilde X^I(x) =X^I_i(x) \chi^i(y) +X^I_a(x) u^a +\underline{X}^I_a(x) v^a\,.
\ea
Here, the fields $X^I_a(x)$ and $\underline{X}^I_a(x) v^a$ are ghost fields.
As we have seen reapeatedly, one may put
\ba
\underline{X}^I_a(x) =0,\quad
X^I_a=\mbox{const}.
\ea
By change of basis in the transverse direction $\mathbf{R}^8$, 
one may put
\ba
&&\tilde X^a(x,y ) =X^a_i(x) \chi^i(y) +\lambda_a u^a \quad (a=1,2,3)\,,\nt
&&\tilde X^I(x, y)=X^I_i(x) \chi^i(y) \quad (I=4,\cdots,8)\,.
\ea
where $\lambda_a$ are constant numbers.
This is exactly the assignment by which we can
reproduce the M5-brane action from BLG model
(for example, eq.\,(30) in \cite{HM}).
Various kinetic terms on M5 world volume 
such as $(\partial_{y^a} X^i)^2$ ($i=4,\cdots,8$) are generated 
from the extra term in $\tilde X^a$.
All the other analysis in \cite{HM, HIMS} remain the same
and we have the same conclusion.

We note that if we do not include these extra terms, the
BLG model associated with this 3-algebra would contain
infinite number of massless mode even when we 
consider the compactification on $T^3$.  On the other hand,
if we use the M5 action in \cite{HM, HIMS} on $T^3$,
we can produce the Kaluza-Klein mass correctly since
we have the kinetic term as mentioned.
Therefore, the generation of Kaluza-Klein mass on M5
can be again reduced to the inclusion of pairs of Lorentzian
norm generators in the Nambu-Poisson 3-algebra.

\section{Conclusion and Discussion}

In this paper, we considered some generalizations of the Lorentzian
Lie 3-algebras and studied the BLG models based on the
symmetry.  In the examples we studied, we naturally
obtain the string/M theory compactifiction on the torus.
The mass term generated by the Higgs fields
can be identified with the Kaluza-Klein mass 
in the toroidal compactification.
The dimension of the torus can be identified with the
number of negative-norm generators of the 3-algebra.
We also argued that one may use our technique to
consider the D-brane system where its gauge symmetry
is described by infinite dimensional loop algebras. 

We do not believe that our examples exhaust
all possible 3-algebras which are relevant to M/string
theories. For example, we did not fully examine the infinite 
dimensional case with $F^{ijkl}\neq 0$. 
Another interesting possibility is the description of
more general background, such as orbifolds, through
different choices of Lorentzian Lie (3-)algebras.

\subsection*{Acknowledgments}

We appreciate partial financial support from
Japan-Taiwan Joint Research Program
provided by Interchange Association (Japan)
by which this collaboration is made possible.

The authors thank Yosuke Imamura and Ta-sheng Tai for helpful discussions. 
Y. M. would like to thank the hospitality of the string theory
group in Taiwan.
The work of P.-M. H. is supported in part by
the National Science Council,
and the National Center for Theoretical Sciences, Taiwan, R.O.C. 
Y. M. is partially supported by
KAKENHI (\#20540253) from MEXT, Japan.
S. S. is partially supported by Global COE Program 
``the Physical Sciences Frontier'', MEXT, Japan.


\end{document}